\documentclass[journal,twoside,web]{ieeecolor}
\usepackage{tmi}
\usepackage{cite}
\usepackage{amsmath,amssymb,amsfonts}
\usepackage{pifont}
\usepackage[ruled,linesnumbered]{algorithm2e}
\usepackage{graphicx}
\usepackage{textcomp}
\usepackage{epsfig}
\usepackage{graphicx}
\usepackage{amsmath}
\usepackage{amssymb}
\usepackage{multirow}
\usepackage{booktabs,caption}
\usepackage[flushleft]{threeparttable}
\usepackage{optidef}

\usepackage{graphicx}
\usepackage{xcolor}
\usepackage{hyperref}
\hypersetup{colorlinks,allcolors=black}

\definecolor{dark}{rgb}{0.3,0.3,0.3}
\definecolor{light}{rgb}{0.75,0.75,0.75}
\definecolor{white}{rgb}{1,1,1}

\DeclareRobustCommand{\legendsquare}[1]{%
  \textcolor{#1}{\rule{1ex}{1ex}}%
}

\newcommand{\sqboxs}{1.0ex}
\newcommand{\sqboxf}{0.3pt}

\newcommand{\sqboxEmpty}[1]{%
  \begingroup
  \setlength{\fboxrule}{\sqboxf}%
  \setlength{\fboxsep}{-\fboxrule}%
  \textcolor{#1}{\fbox{\rule{0pt}{\sqboxs}\rule{\sqboxs}{0pt}}}%
  \endgroup
}

\DeclareMathOperator*{\argminA}{arg\,min}

\def\BibTeX{{\rm B\kern-.05em{\sc i\kern-.025em b}\kern-.08em
    T\kern-.1667em\lower.7ex\hbox{E}\kern-.125emX}}
\begin{document}

\title{Optimal MRI Undersampling Patterns for Ultimate Benefit of Medical Vision Tasks}

\author{Artem Razumov, Oleg Y. Rogov, and Dmitry V. Dylov
\thanks{This work was supported by the Skoltech-MIT NGP Program (Skoltech-MIT joint project).}
\thanks{Authors are with 
the Skolkovo Institute of Science and Technology
Moscow, Russia (corresponding author e-mail: d.dylov@skoltech.ru).}}

\maketitle

\begin{abstract}
To accelerate MRI, the field of compressed sensing is traditionally concerned with optimizing the image quality after a partial undersampling of the measurable \textit{k}-space. 
In our work, we propose to change the focus from the quality of the reconstructed image to the quality of the downstream image analysis outcome. 
Specifically, we propose to optimize the patterns according to how well a sought-after pathology could be detected or localized in the reconstructed images. 
We find the optimal undersampling patterns in \textit{k}-space that maximize target value functions of interest in commonplace medical vision problems (reconstruction, segmentation, and classification) and propose a new iterative gradient sampling routine universally suitable for these tasks. 
We validate the proposed MRI acceleration paradigm on three classical medical datasets, demonstrating a noticeable improvement of the target metrics at the high acceleration factors (for the segmentation problem at $\times$16 acceleration, we report up to 12\% improvement in Dice score over the other undersampling patterns).
\end{abstract}


\section{Introduction}
With its excellent soft-tissue contrast and the absence of ionizing radiation, the value of Magnetic Resonance Imaging (MRI) in modern healthcare cannot be overstated. Yet, even today, this modality remains known for its long data acquisition times, requiring anywhere from 15 to 60 minutes to complete a single scan.
Efficient ways to accelerate the imaging process include the use of contrasting chemicals\cite{Debatin1998}, improved magnetic field gradients\cite{Prakkamakul2016,Tsao2010}, intricate excitation pulse sequences\cite{Bernstein2004,Middione2020}, and parallel imaging approaches\cite{sodickson1997simultaneous,pruessmann1999sense,Griswold2002}.

In lieu of the material-based and the hardware-based methods to accelerate MRI, another perspective to the challenge is offered by the field of compressed sensing\cite{Debatin1998,candes-cs,Lustig2007}. 
The methodology entails an incomplete sampling of the raw \textit{k}-space data that an MR machine can acquire, with the consequent digital compensation for the artifacts caused by the undersampling.
To comply with the Nyquist-Shannon-Kotelnikov sampling theorem and to eliminate the appearance of the aliasing effects\cite{ye2019compressed}, the reconstruction models require incorporating additional \textit{a priori} knowledge about the data (e.g., via total variation~\cite{TV}, heuristic\cite{zijlstra2016evaluation}, and other regularization methods\cite{FastITSA, akccakaya2011low}).
 
The latter line of research has witnessed a recent explosive growth of interest following the publication of the open \texttt{fastMRI} dataset\cite{zbontar2019fastmri,FastMRIGAN2020TCI ,newfastMRI2020REFatTMI}. The deep learning community eagerly unholstered the arsenal of available reconstruction methods, ranging from basic U-Net-like models\cite{Knoll_2020}, to those that incorporate specifics of the imaging process \cite{Pezzotti2020, Ramzi2020}, to superresolution methods \cite{belov2021ultrafast}. The deep learning models proved superior to the classical reconstruction models in compressed sensing, ultimately resulting in their adoption by the MR industry~\cite{DLinMRImachines}.

Classical or deep-learning-based, the compressed sensing models proposed thus far have all had one optimization target at their core: \textit{the quality of the reconstructed image}\footnote{gauged either quantitatively by a common metric, such as SSIM\cite{SSIMloss}, or by a perception of the image by the radiologists \cite{belov2021ultrafast}.}. Although logical and intuitive, such an approach dismisses what happens to the reconstructed images further down the line -- in the downstream image analysis and during the final decision-making. In our work, we aspired to reconsider the MRI \textit{k}-space undersampling problem from the standpoint of the effect that the undersampling has on the ultimate quality of the outcome in medical vision tasks, such as those that are embedded in the diagnostic decision support systems. 

Specifically, we are motivated to find those undersampling patterns in \textit{k}-space that optimize how well a sought-after pathology or an anatomic object could be detected or localized in the reconstructed images (\textit{e.g.}, the heart chambers in the popular \texttt{ACDC} dataset\cite{ACDC} or the brain tumor areas in \texttt{BraTS}\cite{BRATS3}).
We find the optimal undersampling patterns in \textit{k}-space that maximize a target value functions of interest in commonplace medical vision problems (reconstruction, segmentation, and classification) and propose a new iterative gradient sampling routine universally suitable for these tasks. It does not matter if the undersampling pattern ruins the look of the reconstructed image \textit{as long as it improves the target value function} (\textit{e.g.}, Dice score for segmentation problem). What undersampling patterns can accomplish that?

\subsection{Prior Work}

Approaches to accelerate MRI represent one the most actively developing areas of the recent years~\cite{Pineda2020}, with compressed sensing\cite{Liang2010} and parallel imaging~\cite{Deshmane2012} techniques covering a large cohort of publications. The other methods include dictionary-learning algorithms~\cite{Lu2016}, advanced total variation~\cite{Raja2014} and tensor~\cite{fMRIpatchTensor} methods, cross-sampling strategies~\cite{2dCrossSampling}, recurrent inference machines\cite{Putzky}, and others. In these reconstruction-focused works, the optimization of the ultimate value function of interest to a clinical application, such as a Dice score, has been dismissed.

As mentioned above, the undersampling problem has experienced a resurgence with the advent of deep learning~\cite{Knoll_2020, Pezzotti2020, Ramzi2020,bahadir2020deep,xuan2020learning,BakkerHW20,Darestani2021measuring,brain_mri_time,kDL_acc_MRI} and the publication of the \texttt{fastMRI} benchmarks\cite{zbontar2019fastmri,newfastMRI2020REFatTMI}. The most recent works report the adaptation of deep learning methods to \textit{k}-space~\cite{kDL_acc_MRI} and the use of image-to-image translation / superresolution models~\cite{belov2021ultrafast}.

Of particular relevance to the results presented herein are the works that search for the optimal mask via the acquisition trajectory design \cite{liu2012under}, b-spline parametrization~\cite{bjork}, and the works\cite{Zhu2018,Aggarwal2019,Eo2018} that consider the value of using deep-learning algorithms on undersampled data for the MRI image processing. Noteworthy, these works either consider simplistic patterns (such as, central-weighted or the fixed \texttt{fastMRI} masks with, at best, a randomized side-band selection) or separate the reconstruction and the image analysis optimization problems into separate routines in their pipelines.

Considered in this work are three popular datasets: \texttt{fastMRI}~\cite{zbontar2019fastmri}, \texttt{ACDC}~\cite{ACDC}, and \texttt{BraTS}~\cite{BRATS1}. In-depth reviews~\cite{BRATS2} and \cite{Muckley2020} describe validation of MRI acceleration methods on \texttt{BraTS} and \texttt{FastMRI} datasets; and the review \cite{Seo2019} covers state-of-the-art (SOTA) for the \texttt{ACDC} data. 

\subsection{Contributions}
The contribution of this paper is in the following:

\begin{enumerate}
\item We propose a new paradigm for accelerating MRI, where the undersampling of k-space occurs intelligently for the benefit of the downstream image analysis. 
\item In this paradigm, it does not matter if the learned pattern ruins the look of the image; what matters is that it boosts the values of the target metrics.
\item A new iterative gradient sampling (IGS) algorithm learns optimal undersampling patterns for specific medical vision tasks in a given application (cardiac, neurological, and orthopedic utility confirmed).
\item We report noticeable improvement to target value metrics compared to the other undersampling patterns, such as \texttt{fastMRI} or central masks. 
\item The method proves especially instrumental at the highest acceleration factors when a diagnostically valuable image quality becomes impossible to obtain.
\end{enumerate}


\begin{figure*}[t]
    \centering
    \includegraphics[width=0.9\textwidth]{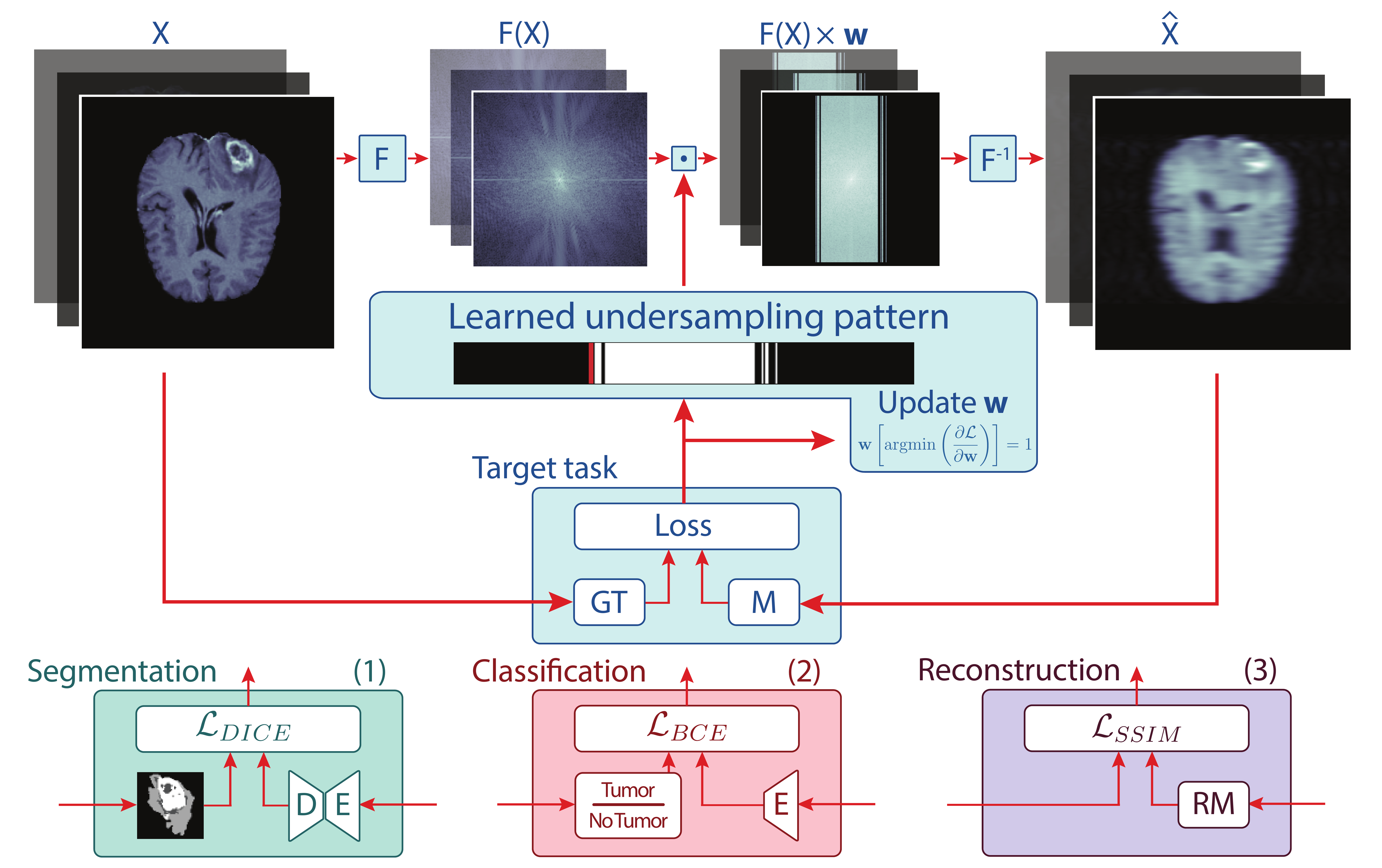} 
    \caption{Framework of the proposed pipeline to learn optimal undersampling patterns for a given medical vision task. \emph{GT} is the ground truth, \emph{M} is a baseline model (\emph{RM} is the reconstruction model), \emph{D} and \emph{E} stand for decoder for encoder, respectively. In (3), RM could be omitted; then, the reconstruction module will simply find an optimal pattern for the dataset at a fixed acceleration factor.}
    \label{fig:masterfigure}
\end{figure*}

\section{Methods}
In this section, we first formulate the MRI acceleration problem via \textit{k}-space undersampling of a generic pattern. Then, we introduce our strategy to optimize the undersampling pattern for a given medical vision task (segmentation, classification, and reconstruction are considered) by a new iterative gradients sampling approach (IGS). Regardless of the medical vision task at hand, the proposed IGS framework follows the optimization routine shown in Fig.~\ref{fig:masterfigure}.

\subsection{Notation}
Throughout the paper, we assume that we have a dataset $\left\{\left(\boldsymbol{X}, \boldsymbol{Y}\right)\right\}^{n},~n\in\mathbb{N}$.
Let us denote unconstrained fully-resolved images of dimensions $N\times N$ as $\boldsymbol{X}^{N\times N}\in \mathbb{R}^{d}$, undersampling binary patterns for \textit{k}-space as $\textbf{w}\in\left \{ 0, 1 \right \}$, and binary ground truth labels in image space as $\boldsymbol{Y}\in\left \{ 0, 1 \right \}$ (a classification label or a segmentation mask). $\mathcal{F}\in\mathbb{C}^{d\times d}$ and $\mathcal{F}^{-1}$ are the fast Fourier (FFT) and the inverse Fourier transform matrices (iFFT), $\alpha={N_s}/{N}$ is the acceleration factor, where $N_s$ and $N$ are the numbers of \textit{k}-space lines sampled in the partially-filled and in the fully-filled scenarios respectively; $\left \| \cdot  \right \|_{1}$ and $\left \| \cdot  \right \|_{2}$ denote L1 and L2 norms, respectively. Lastly, a hat (\textit{e.g.}, $\hat{X}$ or $\hat{Y}$) indicates a predicted image/label and $S(\cdot)$ denotes the prediction model.

\subsection{Problem}
\noindent We consider the following optimization problem:
\begin{argmini}|s|
  {\textbf{w}}{\mathbb{E}_{\boldsymbol{X}} \left\| (S(X),Y))\right\|}{}{},
\label{eq:problem}
\end{argmini}
where $\left \| \cdot  \right \|$ is a differentiable loss function determined by a specific medical vision task. A stochastic gradient descent (SGD)\cite{GoodfellowDL} for the convolutional neural network (CNN) is commonly employed to solving (\ref{eq:problem}). The common medical image analysis problems considered herein are segmentation, classification, and image reconstruction from the raw undersampled \textit{k}-space data. Therefore, let us explicitly define the target loss functions of interest to these tasks.
\noindent\subsubsection*{(1) Segmentation}
\label{s:prob-segm}
For the medical image segmentation, the Dice loss function is a standard starting point:
\begin{equation}
    \mathcal{L}_{DICE}(Y, \hat{Y})=1-\frac{2 Y \hat{Y}+1}{Y+\hat{Y}+1}
\end{equation}

\noindent\subsubsection*{(2) Classification}\label{s:prob-class}
In the binary classification task, the cross-entropy loss~\cite{GoodfellowDL,LiangSoftmaxOrigin} is used. For each class, given the ground truth score $Y_i$, the loss is defined as:
\begin{equation}
    \mathcal{L}_{BCE} = -\sum_{i=1}^{C=2}Y_{i} \operatorname{log} (\hat{Y}_{i}).
\end{equation}

\noindent\subsubsection*{(3) Reconstruction}\label{s:prob-recon}
For image reconstruction task, we can indeed guide the search for the optimal mask using the output's similarity to the fully-resolved image either by L1 loss
\begin{equation}
    \mathcal{L}_{L1} = || X_{i} - \hat{X_{i}}||_{1},
\end{equation}
or by structural similarity (SSIM) loss:
\begin{equation}
    \mathcal{L}_{SSIM} = -\frac{(2 \mu_{X_i} \mu_{\hat{X}_i} + c_1) \cdot (2 \sigma_{\hat{X}_i X_i} + c_2) }{(\mu_{X_i}^2 + \mu_{\hat{X}_i}^2 + c_1) \cdot (\sigma_{X_i}^2 + \sigma_{\hat{X}_i}^2 + c_2)},
\end{equation}
where $c_{1,2}$ are some coefficients and $\mu_{X_i}$, $\sigma_{X_i}$, $\sigma_{X_i \hat{X_i}}$ are the mean, the variance, and the covariance of the pixel intensities. 

Thus, we are looking for the pattern $\textbf{w}$ for a fixed acceleration factor $\alpha={N_s}/{N}$ that solves (\ref{eq:problem}). Such an undersampling pattern will yield a so-called zero-filled reconstruction~\cite{Bernstein2004} with a specific optimization target set by Eqns. (2)--(5), depending on a medical vision task at hand. For example, when a fixed acceleration of $\times16$ is set, what would be an optimal \textit{k}-space pattern $\textbf{w}$ if our ultimate task is, say, segmentation? We solve this through the iterative sampling approach, proposed in the following section.

\subsection{Iterative gradients approach}
The basic idea of the proposed algorithm is to increment the undersampling pattern sequentially from the zero pattern.
We initialize the undersampling pattern $\textbf{w}$ with a null pattern and put 1 in the center $\left(\textbf{w}[\frac{N}{2}] = 1\right)$, because the gradients of the null pattern will be zero.

At each iteration, we pass all undersampled images through the baseline model and accumulate gradients of the undersampling pattern for all data. 
We perform the FFT of the image $X$, multiply the FFT of the image by the undersampled pattern $\textbf{w}$, and perform an iFFT of the undersampled \textit{k}-space to obtain the undersampled image $\hat{X}$.
\begin{equation}
    \hat{X} = \mathcal{F}^{-1}(\mathcal{F}(X) \cdot \textbf{w}).
\end{equation}
Then, we send the undersampled image $\hat{X}$ into the model $S(\cdot)$, which generates the prediction $\hat{Y} = S(\hat{X})$. Finally, we estimate the value of the target loss function $\mathcal{L} = \mathcal{L}(Y, \hat{Y})$ for this pattern and compute its gradient with respect to the undersampled pattern.

Then, we select the position $i$ of the undersampled pattern with the highest negative value and update the undersampled pattern by putting 1 in the corresponding position ($\textbf{w}[i] = 1$):

\begin{equation}
    i \gets  \argminA_{i \mid \textbf{w}[i] = 0} \frac{\partial \mathcal{L}}{\partial \textbf{w}}\,.
\end{equation}\label{eq:argmin}

Naturally, we want to use the gradient descent on the undersampled pattern to train it. 
But the undersampled pattern is binary, which cannot be trained by the gradient descent. 
Unlike~\cite{xuan2020learning}, where the authors proposed to train the estimation of each position in the undersampled vector, we propose to iteratively estimate the gradients $\frac{\partial \mathcal{L}}{\partial W_{i}}$ of the weights $W$ and add gradients with some coefficient $\eta$ to the trained weights.
In other words, in the parameter space, we estimate the direction to the optimal point, do a small shift in the direction of the gradient, and then estimate the direction to the optimal point again:
\begin{equation}
    \label{eq:graddesc}
    W_{i+1} = W_{i} - \eta \cdot \frac{\partial \mathcal{L}}{\partial W_{i}}.
\end{equation}
As such, this routine imitates the idea of the gradient descent but becomes functional for a binary pattern.

In the end, we are interested in finding such an optimal undersampling vector $\textbf{w}$ that would boost the target metric and would have the minimum possible zero-norm $\min _{\mathbf{w}}\|\mathbf{w}\|_{0}$, such that $\|\mathbf{w}\|_{0}=\operatorname{card}\left\{w_{i} \mid w_{i} \neq 0\right\}$.

\begin{algorithm}[t]
\label{alg:IGS_algorithm}
    \caption{Iterative gradients sampling}
    \SetAlgoLined
    \KwData{
    $X$ - MR images,
    $Y$ - ground truth, 
    $S(\cdot)$ - CNN model, 
    $\mathcal{L}_{target}(\cdot, \cdot)$ - target loss function, 
    $N$ - full sampling size, 
    $N_s$ - partial sampling size, 
    ~~~$\textbf{w}$ - \textit{k}-space undersampling pattern.}
    \KwIn{$X, Y, N, N_s$}
    \KwOut{$\textbf{w}$}
    $\textbf{w} \gets 0$\;
    $\textbf{w}[\frac{N}{2}] \gets 1$\;
    \For{$n \gets$ 0 \KwTo $N_s$}{
        $\hat{X} \gets \mathcal{F}^{-1}(\mathcal{F}(X) \cdot \textbf{w})$\;
        $\hat{Y} \gets S(\hat{X})$\;
        $\mathcal{L} \gets \mathcal{L}_{target}(Y, \hat{Y})$\;
        $i \gets \underset{i \mid \textbf{w}[i] = 0}{\argminA(\frac{\partial \mathcal{L}}{\partial \textbf{w}})}$\;
        $\textbf{w}[i] \gets 1$
    }
    \KwResult{Optimized sampling pattern $\textbf{w}$ for acceleration factor $\alpha={N_s}/{N}$ with respect to model $S(\cdot)$.}
\end{algorithm}

It is instrumental to consider what happens if we omit the constraint on the smallest zero-norm. Then, we know that the best solution is a vector of ones $\textbf{w}_N$, and the worst solution is a vector of zeros $\textbf{w}_0$. And, naturally, there is some optimal transition sequence $\mathcal{W} = (\textbf{w}_0, \textbf{w}_1, \ldots, \textbf{w}_i, \textbf{w}_{i+1},\ldots \textbf{w}_N)$ from the worst solution to the best one, such that each transition $i$ (\textit{i.e.}, increasing the zero-norm of the undersampled vector by one) gives the best prediction scores for a given sampling vector $\textbf{w}_i$. Thus, had we known the optimal transition sequence, we could have simply chosen the number of transitions $N_s$ for the desired quality of the target vision task set by Eqns. (2)--(5).

The proposed algorithm is inspired by the iterative design from~\cite{zijlstra2016evaluation} and iteratively selects \textit{k}-space locations corresponding to the highest errors.

We propose to use the gradients of the sampling pattern $\frac{\partial \mathcal{L}}{\partial \textbf{w}_{i}}$ to find the optimal transitions. For the gradient descent, we want to subtract from the sampling vector its gradient with some coefficient. And in this case, the sampling vector gets the largest value where the gradient has the largest negative value. Thus, we propose to search for the optimal transitions by evaluating the gradient of the target loss values with respect to the sampling vector and by finding the index of the largest negative value within the array. We assume that this index is the optimal position of the sampling vector for the next transition (\textit{i.e.}, adding 1 to its position). We, thus, obtain:
\begin{equation}
    \label{eq:maskiter}
    \textbf{w}_{i+1} = \textbf{w}_i + \mathbb{I}_{N,N}\left [ \underset{j \mid \textbf{w}_i[j] = 0}{\argminA} \left (\frac{\partial \mathcal{L}}{\partial \textbf{w}_i}  \right )  \right ]
\end{equation}
where $\mathbb{I}_{N,N}$ is the identity matrix, $j$ is the index of the line in \textit{k}-space selected in $i+1$'th iteration. The gradient descend form of Eq.~(\ref{eq:graddesc}) is replaced by Eq.~(\ref{eq:maskiter}) in our algorithm.
The proposed algorithm yields the sequence of undersampling patterns $\mathcal{W}$, optimized for a given dataset $X$ with the ground truth labels $Y$ and for a given prediction model $S(\cdot)$. The routine is summarized in Algorithm 1.

\section{Experiments}\label{s:experiments}
\subsection{Datasets}\label{s:datasets}
Experiments were carried out with publicly available MRI datasets: \texttt{ACDC}~\cite{ACDC}, \texttt{BraTS}~\cite{BRATS1}, and \texttt{FastMRI} \cite{Zbontar2018}.

\subsubsection{ACDC} 
These cardiac images were acquired using two different MRI scanners with different magnetic strengths of 1.5 T and 3.0 T. The short axis slices cover the left ventricle from the base to the apex such that one gets a single image every 5 to 10 mm. A complete cardiac cycle is usually covered by 28 to 40 images with a spatial resolution varying from 1.37 to 1.68 mm$^2$/px~\cite{ACDC}. The training data is composed of 100 subjects with a total of 1188 slices and the test data is composed of 50 subjects with a total of 514 slices.
The goal of the original disease diagnosis challenge is to classify/cluster
the cardiac MRI-scans into one of the five groups. Ground truth segmentation masks for the cardiac cavities and myocardium are provided.

\subsubsection{BraTS} The \texttt{BraTS} multi-modal data comprise MRI scans of the brain, including native (T1), post-Gd-contrast T1-weighted (T1Gd), T2-weighted (T2), and T2 Fluid Attenuated Inversion Recovery (T2-FLAIR) volumes, acquired with different clinical protocols and various scanners. Ground truth segmentation masks for gliomas (brain tumors) are provided. The training dataset contained 258 subjects with a total of 35508 slices, and the validation set contained 111 subjects with a total of 15391 slices.

\subsubsection{FastMRI} We use a subset of original knee single-coil \texttt{FastMRI} dataset \cite{Zbontar2018}, acquired with 3T scanner in a proton density mode. The subset contains 10374 scans for training and 1992 scans for validation. We also use the proposed script to generate the default \texttt{FastMRI} patterns for acceleration.

\subsection{Comparison protocols: intact \textit{vs.} fine-tuned models}\label{s:protocols}

In sections \ref{s:r} and \ref{s:d}, we compare IGS sampling to the \texttt{FastMRI} and center-weighted (symmetric) patterns. Our primary acceleration factor of interest is $\times$16 (for a 256$\times$256 images it implies $N_{S}=$16 lines), with more sophisticated rule-based sampling being hard to use for the standard datasets\footnote{The majority of \texttt{FastMRI} works consider $\times$4 and $\times$8 acceleration factors.}.

For the segmentation task, we use U-Net~\cite{Ronneberger} as the baseline model, and U-Net with attention~\cite{UNET-ATT} as the SOTA model for \texttt{ACDC} dataset and 3D U-Net~\cite{UNET-3D} as the acknowledged benchmark for \texttt{BraTS} dataset.
As the first step for each dataset, we trained all segmentation models on the fully-sampled training data.
After that, we run IGS algorithm to obtain optimized undersampling pattern for each model.
At that, point we test the models on the undersampled validation set using \texttt{FastMRI} and Center sampling patterns.
Then, for each model, we use pretrained weights from the previous step and train them for every undersampling pattern and dataset -- that is, we \textit{fine-tune} each model for the undersampled data. 

Additionally, we study how SSIM-measured image quality and the target metric (Dice) change at different acceleration factors for the U-Net. For that, additional experiments for the \texttt{ACDC} dataset and the U-Net model were conducted at acceleration factors ranging from $\times 2$ to $\times 26$.

For the classification problem, we use ResNet18 encoder~\cite{he2015deep} initiated and fine-tuned similarly to the U-Net models in the segmentation task.

\subsection{Implementation details}
\subsubsection{Network training}\label{s:train}
For training U-Net, U-Net with attention, and U-Net 3D, we used a batch size of 8 and trained the models for $10^6$ iterations using Adam optimizer (with $\beta_1$ = 0.900, $\beta_2$ = 0.999, and learning rate $lr$ = 0.001). 
As the scope of this work was to show the benefit of simultaneously optimizing the undersampling pattern with the segmentation model, we do not perform any search for the optimal architecture of the reconstruction network, and we use the basic U-Net configuration. In the fine-tuning mode, the models are trained an additional equal number of epochs for each investigated sampling pattern but the batch size and the optimizer hyperparameters remain intact. We employ an early stop once we reach the maximum validation score. 

The pipeline was implemented using Pytorch v.1.4. Experiments were conducted on a server running Ubuntu 18.04 (32 GB RAM, 64-bit); the training was done with a single NVIDIA Tesla V100 (32 GB). In all experiments, we use a 5-fold cross-validation and report the mean performance. 

\subsubsection{Data preprocessing} While the scans are provided as 3D images, we process them as a stack of independent 2D images, which are then fed into the networks. However, for validation on \texttt{BraTS} we evaluate Dice scores for 3D images. The scans were normalized to zero mean and 1 standard deviation before being saved as a set of 2D matrices, and re-scaled to 256$\times$256 px. When several modalities were available, all of them were concatenated before being used as input to the network. In our experiments, we used only the additive noise as the augmentation. It is important to note that the raw \textit{k}-space data were used only for the \texttt{fastMRI} dataset (as it was made available therein but not in the other two datasets), and the image metrics were evaluated only in the cropped area (centered, 320$\times$320).

%
%
\section{Results}\label{s:r}
\subsection*{(1) Segmentation}
\begin{figure}[b]
    \centering
    \includegraphics[width=\linewidth]{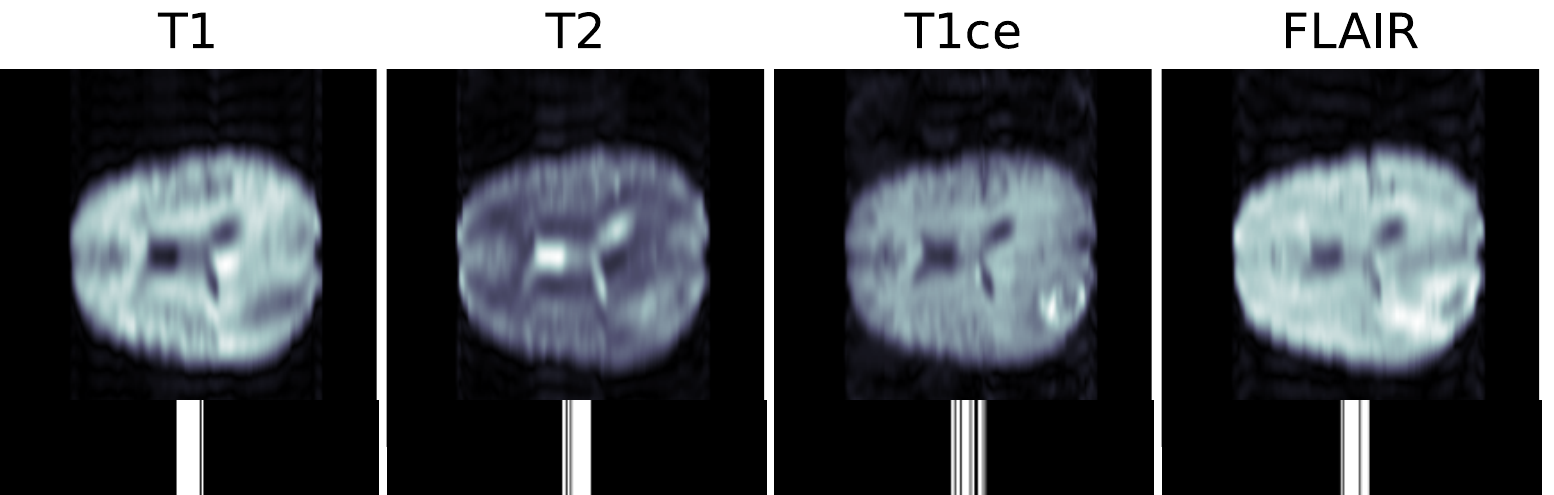}
    \caption{\emph{Top:} \texttt{BraTS} dataset slices for different imaging modes. \emph{Bottom:} corresponding IGS-optimized undersampling patterns at $\times$16 acceleration.}
    \label{fig:mask-example}
\end{figure}

\begin{figure*}[t]
    \centering
    \includegraphics[width=0.9\textwidth]{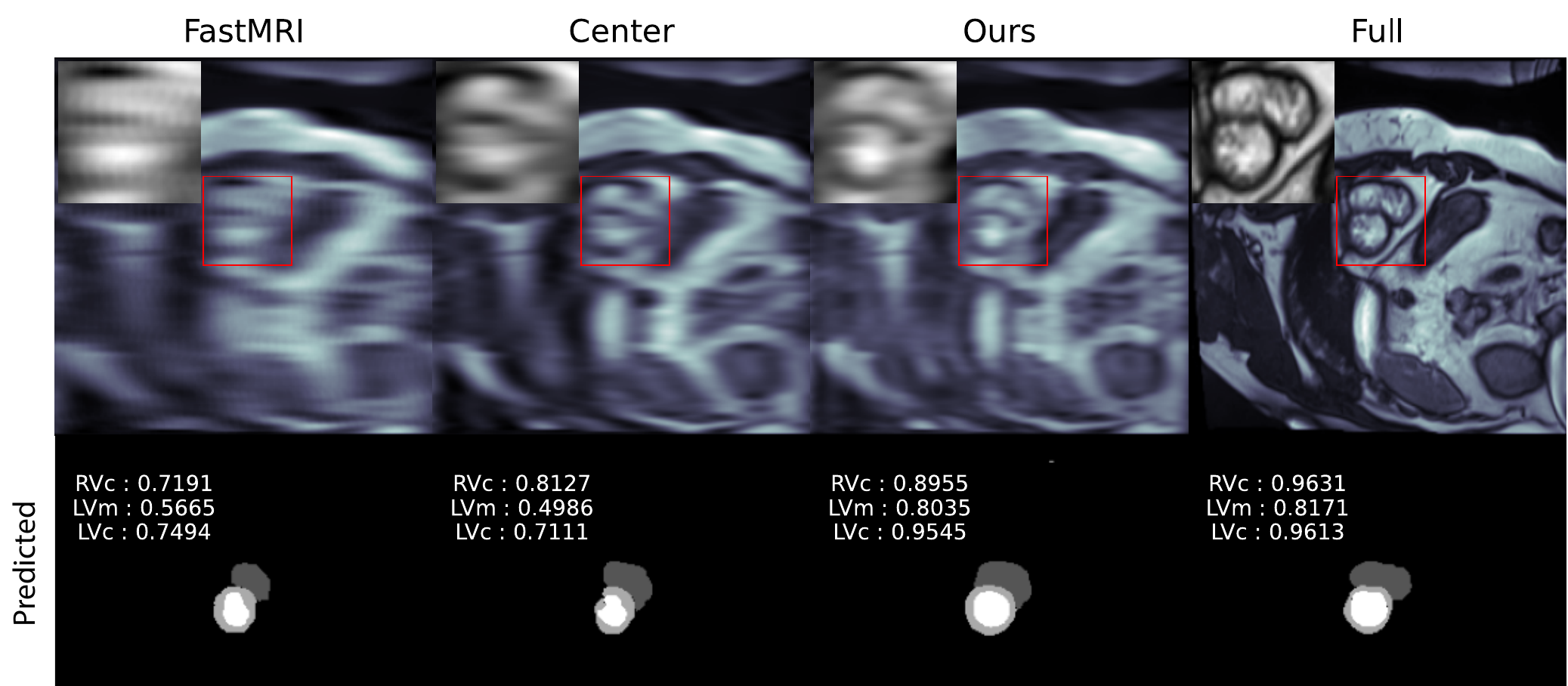}
    \caption{\textbf{Segmentation} results for cardiac structures in the \texttt{ACDC} dataset using fine-tuned U-Net at $\times 16$ acceleration factor: left and right ventricular cavity (LVc \sqboxEmpty{black}, RVc \legendsquare{dark}) and the left ventricular myocardium (LVm \legendsquare{light}). The insets show the magnified areas highlighted in red.}
    \label{fig:ACDC-undersampling}
\end{figure*}

\begin{figure*}[t]
    \centering
    \includegraphics[width=0.9\textwidth]{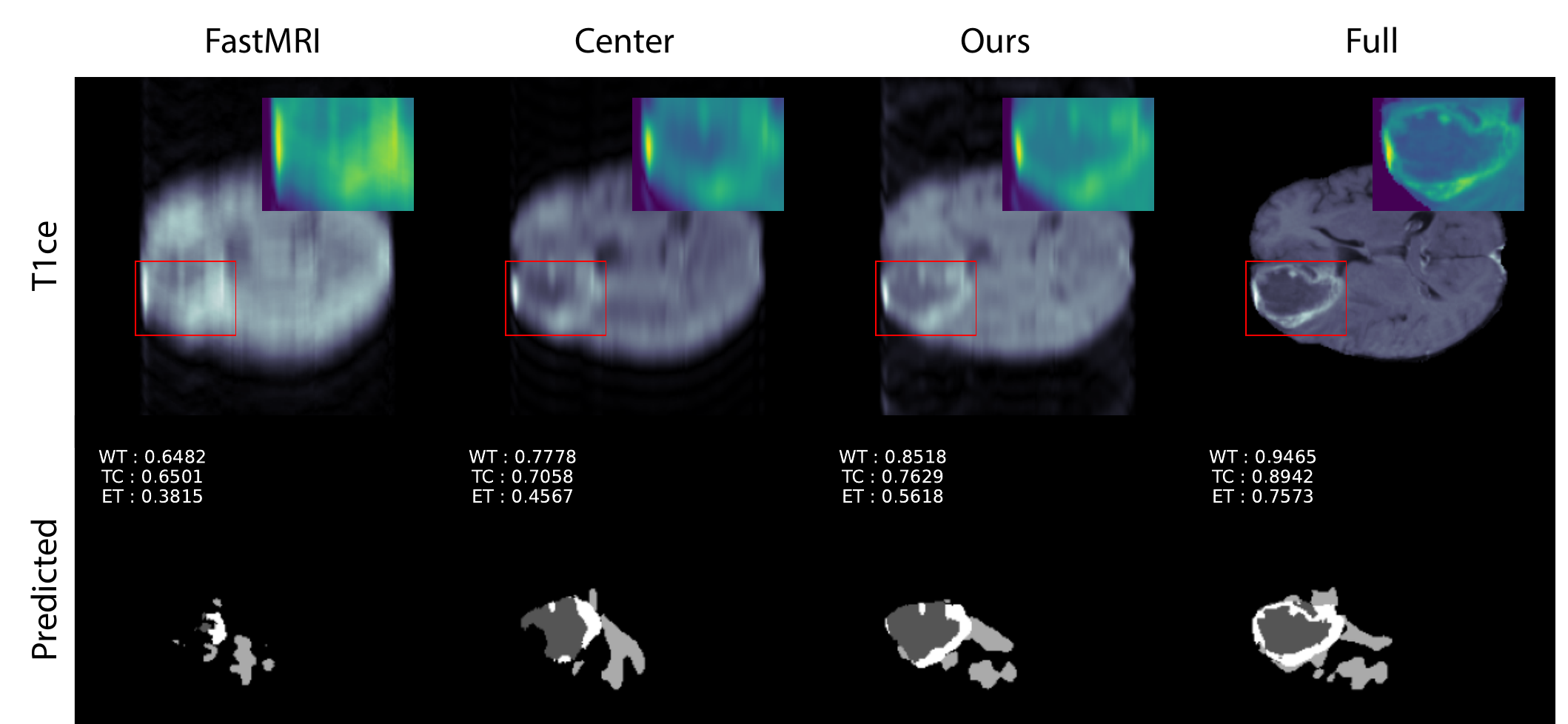}
    \caption{\textbf{Segmentation. }Fine-tuned U-Net undersampling segmentation example with the \texttt{BraTS}2020 dataset at $\times 16$ acceleration factor. Dice scores are provided for the whole tumor (WT \legendsquare{light}), the tumor core (TC \sqboxEmpty{black}), and the enhancing tumor (ET \legendsquare{dark}).}
    \label{fig:BraTS-undersampling}
\end{figure*}

We begin by showing the effect of the learned undersampling pattern on the reconstructed image for various imaging modes (Fig. \ref{fig:mask-example}).
Visual comparison of the segmentation results for different patterns for \texttt{ACDC} dataset is shown in Fig. \ref{fig:ACDC-undersampling} and quantitative comparison is shown in Tables \ref{tab:finetuning_acdc} and \ref{tab:ACDC_acc}. Corresponding results for the \texttt{BraTS} dataset are shown in Fig. \ref{fig:BraTS-undersampling} and in Tables \ref{tab:finetuning_brats} and \ref{tab:BRATS_acc}.

\begin{table}[t]
    \centering
    \begin{threeparttable}

    \caption{\textbf{Segmentation} on \texttt{ACDC} dataset. 5-fold cross-validation Dice scores at $\times$16 acceleration.}
    \begin{tabular}{ccccccc}
    \toprule
    \multicolumn{1}{c}{FT}  & \multicolumn{2}{c}{FastMRI}                           & \multicolumn{2}{c}{Center}   & \multicolumn{2}{c}{\textbf{Ours}}                                 \\ 
    \multicolumn{1}{c}{}                   & \multicolumn{1}{c}{U-Net} & \multicolumn{1}{c}{U-Net-Att} & \multicolumn{1}{c}{U-Net} & \multicolumn{1}{c}{U-Net-Att} & \multicolumn{1}{c}{U-Net} & \multicolumn{1}{c}{U-Net-Att} \\ \midrule 
    \multicolumn{1}{c}{\ding{55}} & 0.658                  & 0.583                      & 0.757                  & 0.734             & \textbf{0.790}                & 0.748                      \\
    \multicolumn{1}{c}{\ding{51}}    & 0.814                  & 0.805                      & 0.857                  & 0.857             & \textbf{0.874}                & 0.871                      \\ \bottomrule
    
        \end{tabular}
    \label{tab:finetuning_acdc}
    \begin{tablenotes}
      \small
      \item \textbf{Note:} results provided without fine-tuning (FT) (\ding{55}) at $p\text{-value} = 0.0136$, with FT (\ding{51}) at $p\text{-value} = 0.0067$.
    \end{tablenotes}
    \end{threeparttable}
\end{table}

\begin{table}[t]
    \centering
    \begin{threeparttable}
    \caption{\textbf{Segmentation} on the \texttt{BraTS} dataset. 5-fold cross-validation Dice scores at $\times$16 acceleration.}
    \begin{tabular}{ccccccc}
    \toprule
    \multicolumn{1}{c}{FT}                   & \multicolumn{2}{c}{FastMRI}                           & \multicolumn{2}{c}{Center}                               & \multicolumn{2}{c}{\textbf{Ours}}                                 \\  
    \multicolumn{1}{c}{}                   & \multicolumn{1}{c}{U-Net} & \multicolumn{1}{c}{U-Net 3D}   & \multicolumn{1}{c}{U-Net} & \multicolumn{1}{c}{U-Net 3D}   & \multicolumn{1}{c}{U-Net} & \multicolumn{1}{c}{U-Net 3D}   \\ \midrule 
    \multicolumn{1}{c}{\ding{55}} & 0.523                  & 0.542                      & 0.618                  & 0.604             & \textbf{0.653}                & 0.602                      \\
    \multicolumn{1}{c}{\ding{51}}    & 0.700                  & 0.642                      & 0.752                  & 0.652             & \textbf{0.773}                & 0.664  \\
    \bottomrule
    \end{tabular}
    \label{tab:finetuning_brats}
    \begin{tablenotes}
      \small
      \item \textbf{Note:} results provided without fine-tuning (FT) (\ding{55}) at $p\text{-value} = 8.3108\times10^{-5}$, with FT (\ding{51}) at $p\text{-value} = 0.0002.$
    \end{tablenotes}
    \end{threeparttable}
\end{table}
\begin{table*}[t]
    \centering
    \caption{\textbf{Segmentation }results. Dice scores on \texttt{ACDC} dataset at $\times$16 acceleration.}
    \begin{tabular}{ccccccccc}
    \toprule
                            & \multicolumn{2}{c}{FastMRI} & \multicolumn{2}{c}{Center} & \multicolumn{2}{c}{Ours}        & \multicolumn{2}{c}{Full} \\ \midrule
                            & U-Net         & U-Net-Att        & U-Net       & U-Net-Att      & U-Net           & U-Net-Att       & U-Net      & U-Net-Att     \\ \midrule
    RVc                  & 0.756        & 0.692           & 0.788      & 0.795         & \textbf{0.813} & 0.790          & 0.868     & 0.899        \\
    LVm                  & 0.675        & 0.638           & 0.716      & 0.727         & \textbf{0.771} & 0.749          & 0.865     & 0.891        \\
    LVc                  & 0.825        & 0.783           & 0.844      & 0.851         & \textbf{0.882} & 0.859          & 0.931     & 0.941        \\ \bottomrule
    \end{tabular}
    \label{tab:ACDC_acc}
\end{table*}
\begin{table*}[t]
    \centering
    \caption{\textbf{Segmentation }results. Dice scores on \texttt{BraTS} dataset at $\times$16 acceleration.}
    \begin{tabular}{ccccccccc}
    \toprule
                            & \multicolumn{2}{c}{FastMRI} & \multicolumn{2}{c}{Center} & \multicolumn{2}{c}{Ours}        & \multicolumn{2}{c}{Full} \\ \midrule
    \multicolumn{1}{l}{} & U-Net         & U-Net 3D          & U-Net       & U-Net 3D        & U-Net           & U-Net 3D         & U-Net      & U-Net 3D       \\ \midrule
    WT                   & 0.808        & 0.739           & 0.821      & 0.739         & \textbf{0.835} & 0.787          & 0.873     & 0.888        \\
    TC                   & 0.640        & 0.650           & 0.673      & 0.623         & 0.690          & \textbf{0.698} & 0.726     & 0.792        \\
    ET                   & 0.387        & 0.448           & 0.442      & 0.390         & 0.468          & \textbf{0.481} & 0.575     & 0.623        \\ \bottomrule
    \end{tabular}
    \label{tab:BRATS_acc}
\end{table*}


\begin{figure}[t]
    \centering
    \includegraphics[width=1\linewidth]{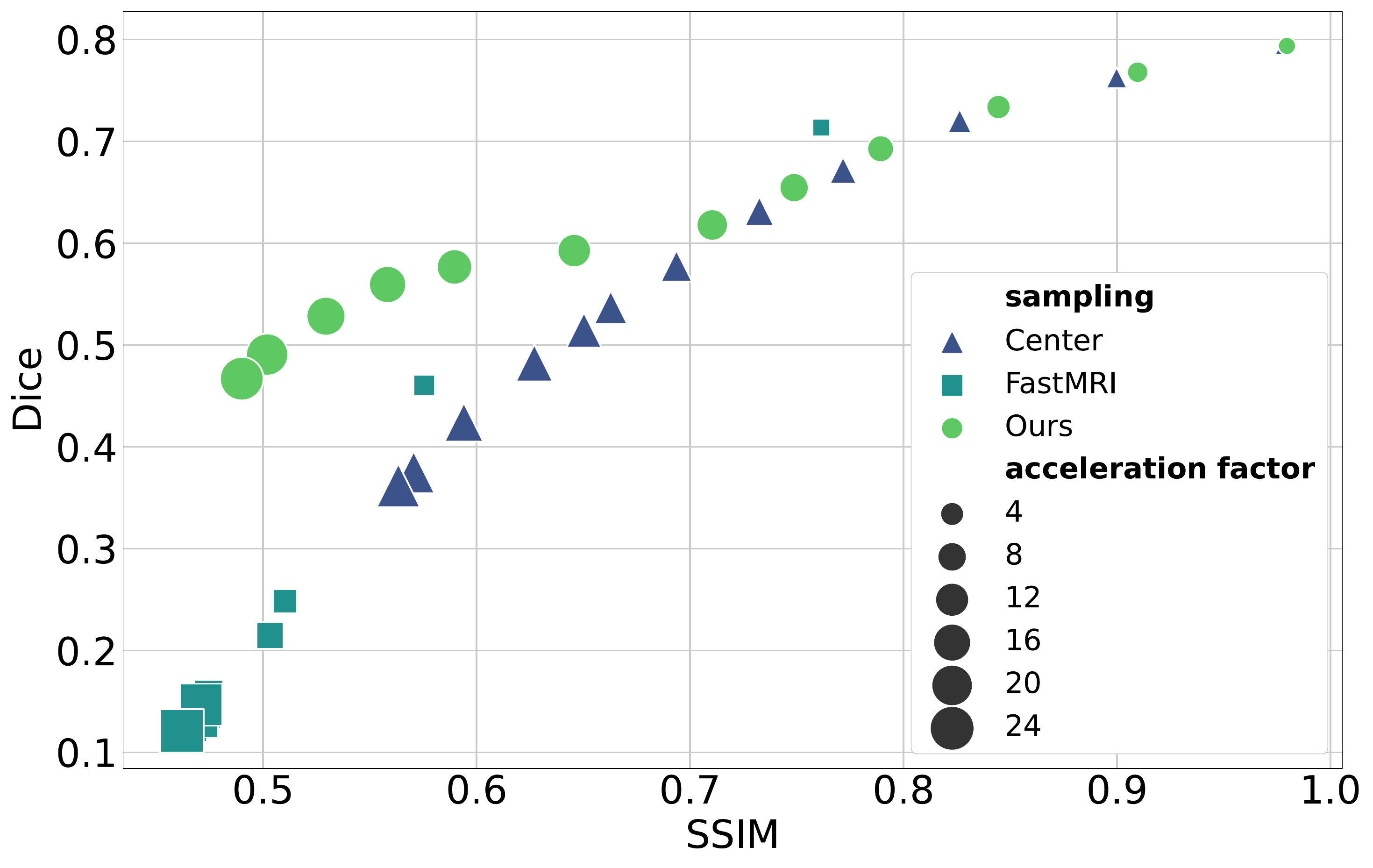}
    \caption{\textbf{Segmentation} \textit{vs.} \textbf{Reconstruction} for \texttt{ACDC} dataset. Measured Dice \textit{vs.} SSIM at different acceleration factors for three undersampling patterns. 
    The size of the mark is proportional to acceleration factor. 
    Note how our sampling boosts segmentation quality even when SSIM drops to the low values at moderate to high acceleration factors.}
        \label{fig:accelerated}
\end{figure}
\begin{figure}[t]
    \centering
    \includegraphics[width=1\linewidth]{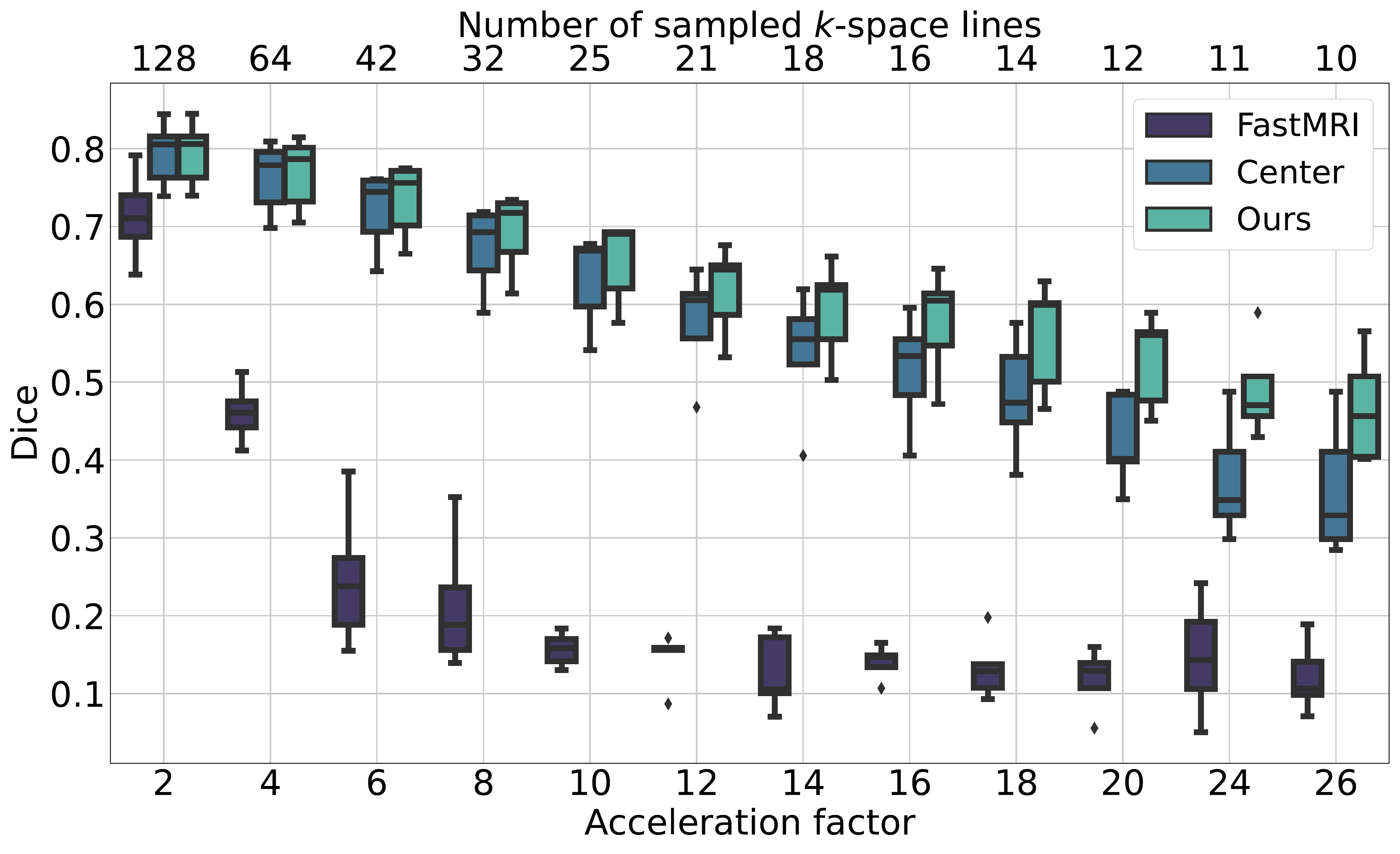}
    \caption{\textbf{Segmentation} performance as a function of acceleration factor for \texttt{ACDC} dataset. Note the growing gap in the Dice scores at acceleration factors beyond $\times 4$.}
        \label{fig:acc-dice-curve} 
\end{figure}

\subsection*{(2) Classification}
\begin{figure}[t]
    \centering
    \includegraphics[width=\linewidth]{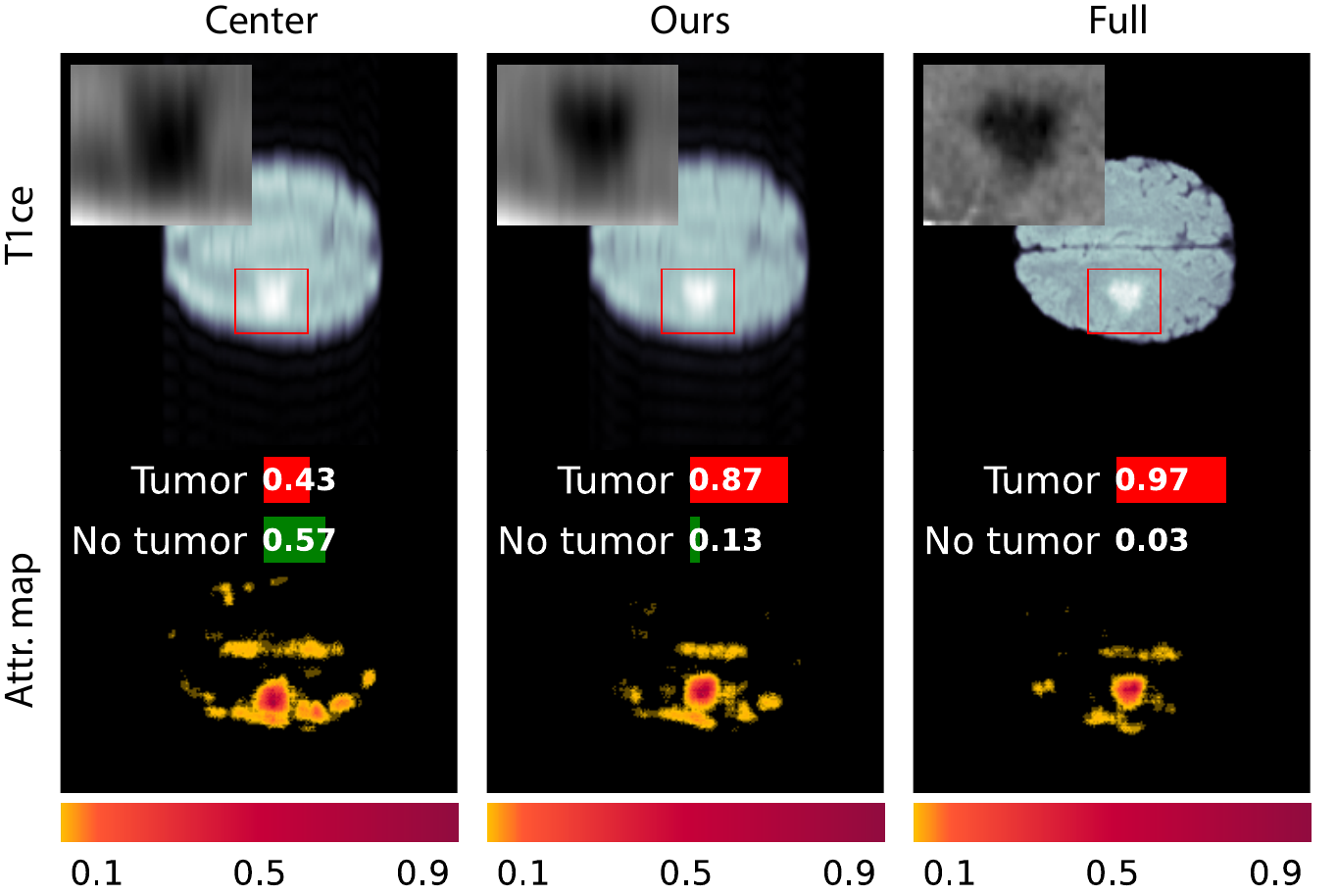}
    \caption{\textbf{Classification } example on \texttt{BraTS} data at $\times 16$ undersampling. Notice how center-weighted pattern fails to classify the pathology while the proposed and the full sampling do it correctly. \emph{Top row}: slices with insets showing the pathology area. \emph{Bottom row}: corresponding attribute maps in pseudo colors. Label predictions are highlighted in red and green.}
    \label{fig:clf-undersampling}
\end{figure}

The classification F1-scores are given in Table~\ref{tab:classification} and the visual results are demonstrated in Fig.\ref{fig:clf-undersampling}, featuring false-colored tumor attribute heatmaps emphasizing which pixel groups have contributed to the classification decision for each undersampling pattern.

\begin{table}[!h]
    \centering
    \caption{\textbf{Classification } results. F1-scores with 5-fold cross-validation on \texttt{BraTS} data at $\times 4$ acceleration.}
    \begin{tabular}{llll}
    \toprule
                 & FastMRI  & Center   & Ours     \\
    \midrule
    CrossEntropy & 0.851 & 0.862 & \textbf{0.873} \\
    Soft Hinge   & 0.834 & 0.848 & \textbf{0.877} \\
    \bottomrule
    \end{tabular}
    \label{tab:classification}
\end{table}

\subsection*{(3) Image reconstruction}
\begin{figure}[t]
    \centering
    \includegraphics[width=\linewidth]{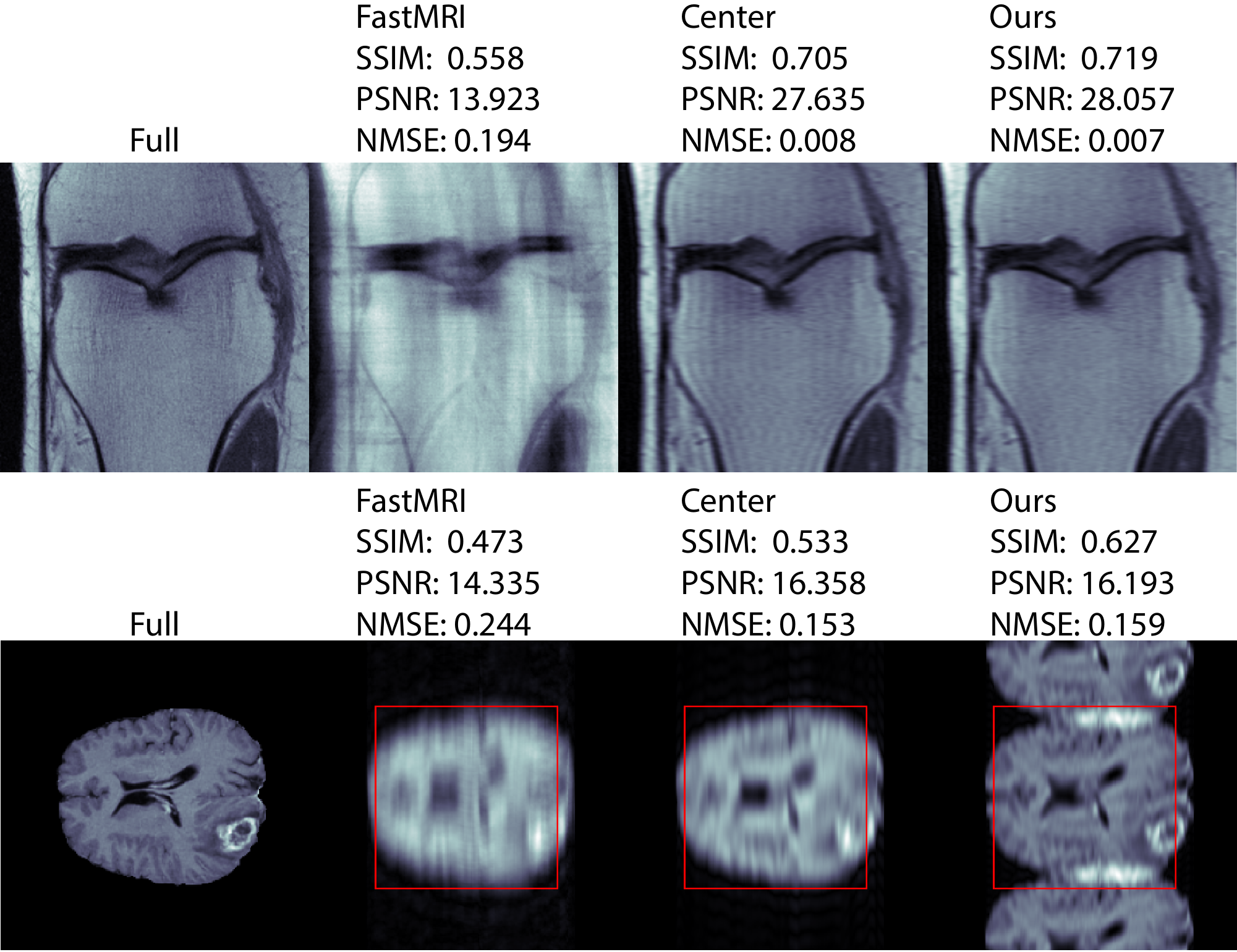}
    \caption{\textbf{Reconstruction. }\textit{Top:} FastMRI zero-filled image reconstruction at $\times 4$ acceleration. Cropped to region of interest. \textit{Bottom:} \texttt{BraTS} zero-filled image reconstruction at $\times 16$ acceleration. The region of interest crop is highlighted with the red rectangle. Relevant metrics are given above the images. Notice evident macro-scale aliasing when one uses our pattern; yet, the routine still maximizes the target metric (SSIM).}
    \label{fig:zf}
\end{figure}

Reconstruction metrics (SSIM, PSNR, NMSE) with omitted RM are provided in Table~\ref{tab:zf-metric}. Samples of the reconstructed images employing the IGS optimization for \texttt{FastMRI} and \texttt{BraTS} datasets are shown in Fig.~\ref{fig:zf}. We report the optimal pattern to maximize a typical reconstruction metric at a fixed acceleration (without studying specific reconstruction models).

\begin{table}[!h]
    \centering
    \caption{\textbf{Reconstruction } results. Measured metrics for \texttt{FastMRI} and \texttt{BraTS} datasets at $\times 4$ acceleration.}
    \begin{tabular}{@{}lllllll@{}}
    \toprule
    & & FastMRI  & &  & BraTS & \\
    \midrule
         Pattern   & SSIM              & PSNR               & NMSE & SSIM              & PSNR               & NMSE              \\ \midrule
    FastMRI & 0.564          & 20.330          & 0.104    & 0.588          & 18.753          & 0.434          \\
    Center  & 0.693          & 26.850          & \textbf{0.029} & 0.635          & \textbf{20.969} & \textbf{0.263} \\
    Ours    & \textbf{0.700} & \textbf{26.861} & \textbf{0.029}      & \textbf{0.678} & 20.354          & 0.268           \\ \bottomrule
    \end{tabular}
    \label{tab:zf-metric}
\end{table}

%
%
\section{Discussion}\label{s:d}
\subsection*{(1) Segmentation}

The results of the cardiac segmentation, shown in Fig. \ref{fig:ACDC-undersampling}, were then assessed for their scaling with regard to acceleration factor.
The resulting 5-fold cross validation outcomes for different undersampling patterns are shown in Figs.~\ref{fig:accelerated} and~\ref{fig:acc-dice-curve}. In Fig.~\ref{fig:acc-dice-curve}, we show how the proposed sampling outperforms the other undersampling patterns at accelerations above $\times 8$ and significantly outperforms them at factors of $\times 16$ and beyond. 

Fig.~\ref{fig:accelerated} illustrates the trade-off between segmentation quality (Dice score) and the image quality that one typically optimizes in the reconstruction stage (SSIM) as the acceleration factor is varied from $\times 2$ to $\times 26$. It is evident that for the standard undersampling, the Dice and the SSIM metrics are almost linear. However, this is not the case for the IGS strategy: as the acceleration becomes extreme ($\times 16$ and above), the Dice drop slows down even though the quality of the reconstructed image is extremely low at this point. 
It should be noted that this is a direct consequence of our optimization routine which maximizes Dice score. 
Center and \texttt{FastMRI} patterns aim to improve the image quality resembling each other in Fig.~\ref{fig:accelerated}. Additionally, \texttt{FastMRI} performs worse at higher acceleration factors due to its pattern sparsity. The deviation amongst the techniques is most noticeable at factors beyond $\times 12$ which is demonstrated in Fig.~\ref{fig:acc-dice-curve} where the sparse \texttt{FastMRI} patterns fail to provide enough detail (Dice scores above 0.2 along with larger variance among the greater Dice values), while here, the Center and the proposed technique show concordant behavior. Although Center patterns show relatively high scores compared to the \texttt{FastMRI}, one can see a clear superiority of the IGS undersampling in the images. We, therefore, conducted further experiments with an acceleration factor of $\times 16$ to study the trade-off.

\subsubsection*{Performance Comparison of Brain segmentation}\label{s:BRATS_res}
As shown in Table~\ref{tab:finetuning_brats}, the IGS sampling outperforms both \texttt{fastMRI} and center patterns. U-Net with IGS samplings shows similar results to \texttt{ACDC} SOTA + IGS, and is significantly better than \texttt{BraTS} SOTA + IGS ($p < 0.0002$).
Notable changes exist also in the patterns for the different modalities at the same acceleration rate $\times 16$  (Fig.~\ref{fig:mask-example}), reinforcing our intuition that the proposed optimization routine should be application- and mode-specific.

As one can observe in Tables~\ref{tab:finetuning_acdc} and~\ref{tab:finetuning_brats}, U-Net with IGS sampling outperforms others sampling and models, except for \texttt{BraTS}: here U-Net 3D with IGS sampling shows better Dice scores for ET and TC tissues, which could be attributed to the way the small and the large volumetric details develop at such acceleration.
Yet, IGS sampling preserves important image features at these scales\footnote{Only those features that maximize the Dice score.}, although it adds more aliasing artifacts than the symmetric central sampling (Fig.~\ref{fig:BraTS-undersampling}). Fragmented small tumor details are preserved well with the proposed technique, while the Center pattern dismisses the tumor core regions as well as the fragments of the whole tumor borders. Similarly, the \texttt{FastMRI} pattern demonstrates inferior segmentation performance at this acceleration.

Despite the visually perceived blur in the undersampled images, the Dice score is maximized for all classes in all datasets. Notice distortions in and outside the region of interest (ROI): more details are preserved inside the ROI; yet, the undersampling-induced artifacts appear in the other (irrelevant) areas of the image. The results of maximizing the Dice scores can be seen in the actual predicted masks in Fig.~\ref{fig:ACDC-undersampling} (bottom). 

Compared to \texttt{FastMRI} and Center patterns, the proposed IGS patterns yielded better anatomical structure similarity to the particular shape of RVc. While optimizing the Dice towards the maximum values, the model led to the appearance of the aliasing at large object scales.

\subsubsection*{Performance Comparison of Cardiac Segmentation}\label{s:ACDC_res}
For \texttt{ACDC} segmentation, U-Net with the proposed sampling outperformed the other sampling, both with and without fine-tuning. Although SOTA models (U-Net with attention) demonstrate better performance on the fully sampled data, the ordinary U-Net shows better results with the IGS-undersampled data. Interestingly, although the U-Net with attention has better metrics than the regular U-Net, it performs worse on the undersampled data. We speculate that the attention on the highly undersampled data limits the capacity of the model.

The effect similar to that described in Section~\ref{s:BRATS_res} is seen in Fig.~\ref{fig:ACDC-undersampling}: the cardiac cavities are hardly distinguishable for both \texttt{FastMRI} and Center patterns, distorting the right ventricle boundary, yet the proposed sampling keeps all the features of the cavities that are vital for the segmentation performance.

\subsection*{(2) Classification}\label{s:class}
To demonstrate the applicability to a classification task, we trained ResNet-18 to classify \texttt{BraTS} with respect to the presence of tumor, splitting the data into two classes (Tumor: 17281 images, No Tumor: 18227 images). Different loss functions proved functional with the IGS-learned undersampling patterns.

To show how attribution maps alter with the sampling (Fig.~\ref{fig:clf-undersampling}) we use the integrated gradients method \cite{sundararajan2017axiomatic}. Particularly, the attribution maps are more localized, and the false-color highlighted tumorous spots are similar to those in the fully-sampled images, while the texture and the grainy structure are lost due to the undersampling. 
Notably, all undersampling patterns washed away the longitudinal cerebral fissure.
Once again, we confirm that it does not matter how the image looks; if the goal is to detect gliomas, the \textit{k}-space undersampling should be performed precisely as determined by the proposed IGS algorithm. 

\subsection*{(3) Image reconstruction}
The proposed IGS approach can be applied to the zero-filled reconstruction task, which answers the most general question: what is the optimal pattern at a given acceleration factor? As shown on the \texttt{BraTS} and \texttt{FastMRI} raw \textit{k}-space data, IGS allows finding patterns that yield the reconstruction quality most similar to the fully-sampled image. 

\subsubsection*{FastMRI} We considered L1-loss, SSIM, and PSNR, as $\mathcal{L}_{target}(\cdot, \cdot)$. The best improvement was observed with the L1-loss. Although there is a little improvement compared to the Center pattern (around +9\%, Table~\ref{tab:zf-metric}), visually one can see that the anatomic details are preserved better via the IGS approach (see Fig.~\ref{fig:zf}, top).
Despite the presence of large-scale aliasing artifacts, the proposed approach yielded higher SSIM and PSNR values and reduced the NMSE. Notice the distortions in the \texttt{FastMRI}-reconstructed images and the reflection artifacts that appeared with the Center pattern.

\subsubsection*{BraTS} For \texttt{BraTS} dataset, IGS with L1-loss replicated the Center pattern outcome; however, with the SSIM loss, there is a significant improvement in SSIM metrics (see Table~\ref{tab:zf-metric}). Following Fig.~\ref{fig:zf} (bottom), one can see how IGS pattern preserves the important image details better even by the naked eye, yet -- as before -- large aliasing artifacts do appear. We note that such artifacts are easy to correct in the post-processing \cite{candes-cs}, but we left them intact for consistency. 

In \texttt{BraTS} images, brain occupies at most 60\% of the field-of-view, while in \texttt{FastMRI} scans, the knee takes about 80-90\% of the frame, which is the reason why the aliasing appears in the former but not in the latter case.
\begin{figure*}[!h]
    \centering
    \includegraphics[width=0.9\textwidth]{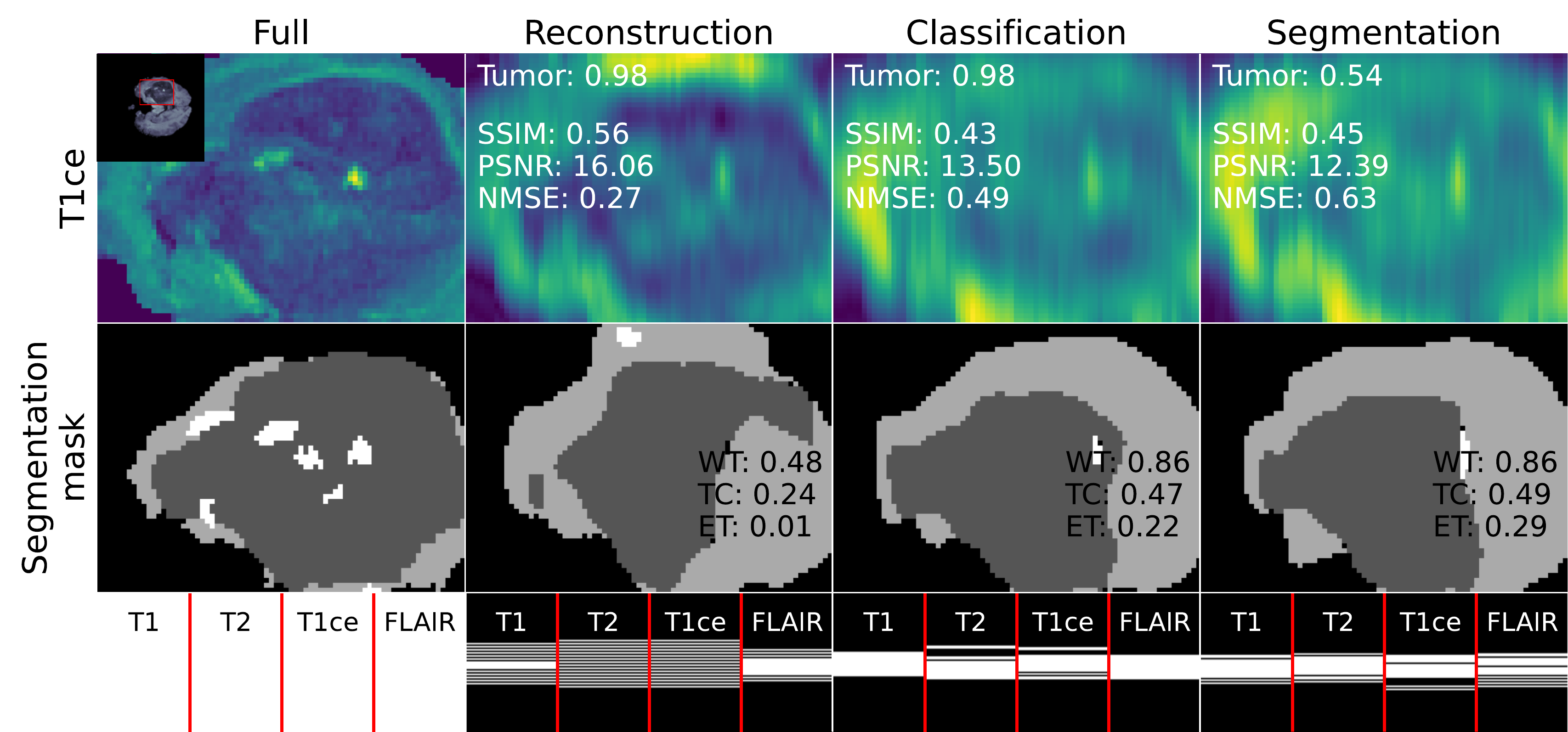}
    \caption{Comparison of IGS-generated patterns and reconstructed images for different medical vision tasks at $\times 16$ acceleration factor. \emph{Top row:} Magnified pathology views from the reconstructed slices. The inset over the fully sampled image is provided to show the location of the pathology. \emph{Middle row:} Post-IGS U-Net performance for various tasks, Dice values overlaid for whole tumor (WT \legendsquare{light}), the tumor core (TC \sqboxEmpty{black}) and the enhancing tumor (ET \legendsquare{dark}). \emph{Bottom row:} IGS sampling patterns are shown for corresponding modalities.}
    \label{fig:gallery}
\end{figure*}

\subsubsection*{Ultimate sampling for a given application} 
Finally, we visually compare the undersampling patterns found by our algorithm for different medical vision objectives (Fig.~\ref{fig:gallery}). Optimizing the reconstruction metrics alone leads to complete loss of the tumor boundaries, as shown in Fig.~\ref{fig:gallery}, middle row. One may also see the dissipation of vertical tumor boundaries with the overall preservation of the shape, which could, perhaps, be improved by devising a 2D variant of the IGS method. 
We find that each IGS-found pattern is optimal for each specific task, allowing to supersede the standard center and \texttt{fastMRI} sampling strategies for these applications.

\section{Conclusions} 
In this work, we proposed a new paradigm for accelerating MRI, where the undersampling of \textit{k}-space occurs intelligently for the ultimate benefit of the downstream image analysis tasks. As we observed, it does not matter if the learned \textit{k}-space pattern ruins the look of the reconstructed image; what matters is that the pattern boosts the values of the target metrics, ultimately holding promise to assist radiologists in detecting and localizing pathologies at low-field or high-speed imaging settings. A new iterative gradient sampling (IGS) algorithm was shown to be capable of finding optimal undersampling patterns in several medical vision tasks and applications, with cardiac, neurological, and orthopedic utility confirmed herein. Remarkably, our intelligent undersampling pattern could boost the target performance score by up to 12\% (Dice metric for the pathology localization problem) over the result obtained with the default \texttt{fastMRI} or with the center-weighted patterns.

In addition to a number of applications considered, we also show that the proposed undersampling method is agnostic of which MRI mode is at use, with applicability confirmed for all imaging regimes and for a broad range of acceleration factors. While we primarily studied the extent of undersampling frequently reported in the other compressed sensing publications (up to $\times 16$), our method proved especially instrumental at the moderate and the highest acceleration factors -- when a diagnostically valuable image quality becomes impossible to obtain. 

These low-quality images are not expected to be useful in typical clinical examinations yet; however, they are already in demand in certain other (non-diagnostic) applications, where the coarse images oftentimes prove sufficient: \textit{e.g.}, surgical planning in the radiological suites, radiation therapy planning in oncology, or uncertainty estimation in decision support systems. We believe these intelligently undersampled task-oriented \textit{k}-space measurements are bound to improve with time because of the growth of the available data, which will gradually pave the way towards true ultrafast MRI of high diagnostic value.

Near-future research directions include adaptation of the undersampling strategy to 2D and 3D~\cite{bjork}, application of the domain adaptation methods to the cross-task undersampling patterns~\cite{LempitskyDomainAdaptation, ProkopenkoAIRTatMICCAIworkshop}, and the study of the compound loss functions to optimize several medical vision objectives at once~\cite{brule_arxiv_Yarik}. We release our code publicly at Github\footnote{https://github.com/cviaai/IGS/}.

\bibliographystyle{IEEEtran}
\bibliography{biblio}
\newpage
\section{Supplementary material}
\subsection{Overview}
This supplementary material is structured as follows. In Section~\ref{s:acdc} we report additional cardiac segmentation examples with the ACDC dataset. In Section~\ref{s:brats} we report additional tumor segmentation examples with the BraTS2020 dataset. For both datasets we provide the results of the 5-fold cross-validation in Section~\ref{s:CV}.

\subsection{Cardiac segmentation examples}
\label{s:acdc}

\begin{figure}[!h]
    \centering
    \includegraphics[width=\linewidth]{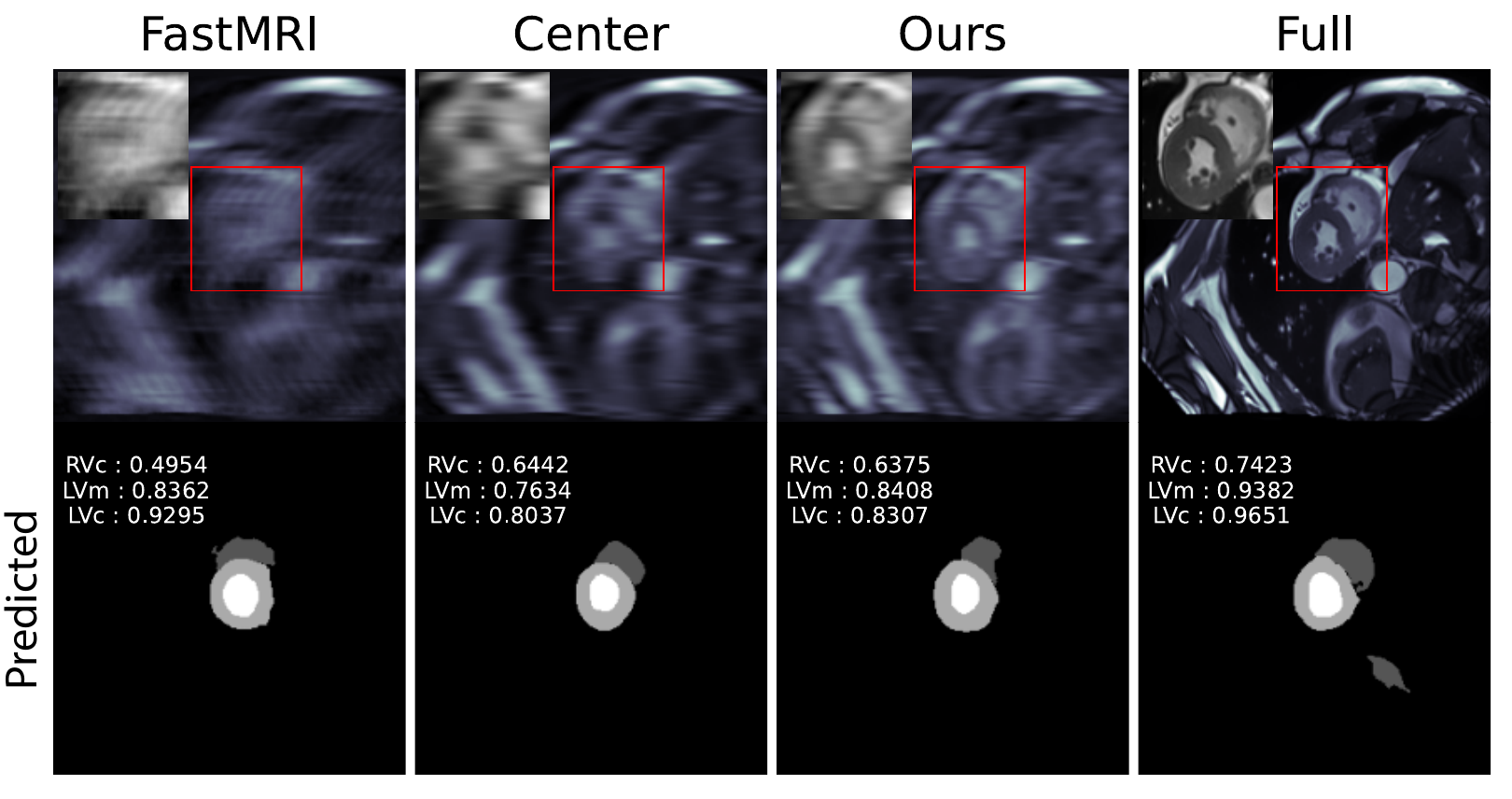}
    \caption{\textbf{Segmentation. }Fine-tuned U-Net undersampling segmentation example on ACDC at $\times$16 acceleration: left and right ventricular cavity (LVc \sqboxEmpty{black}, RVc \legendsquare{dark}) and the left ventricular myocardium (LVm \legendsquare{light}). Undersampling by low-resolution pattern (Center) removes border of ventricles, whereas our pattern better preserve ventricle borders, although Dice score for RV is slightly better with Center pattern.}
\label{sf:unet_acdc_segment}
\end{figure}

\begin{figure}[!h]
    \centering
    \includegraphics[width=\linewidth]{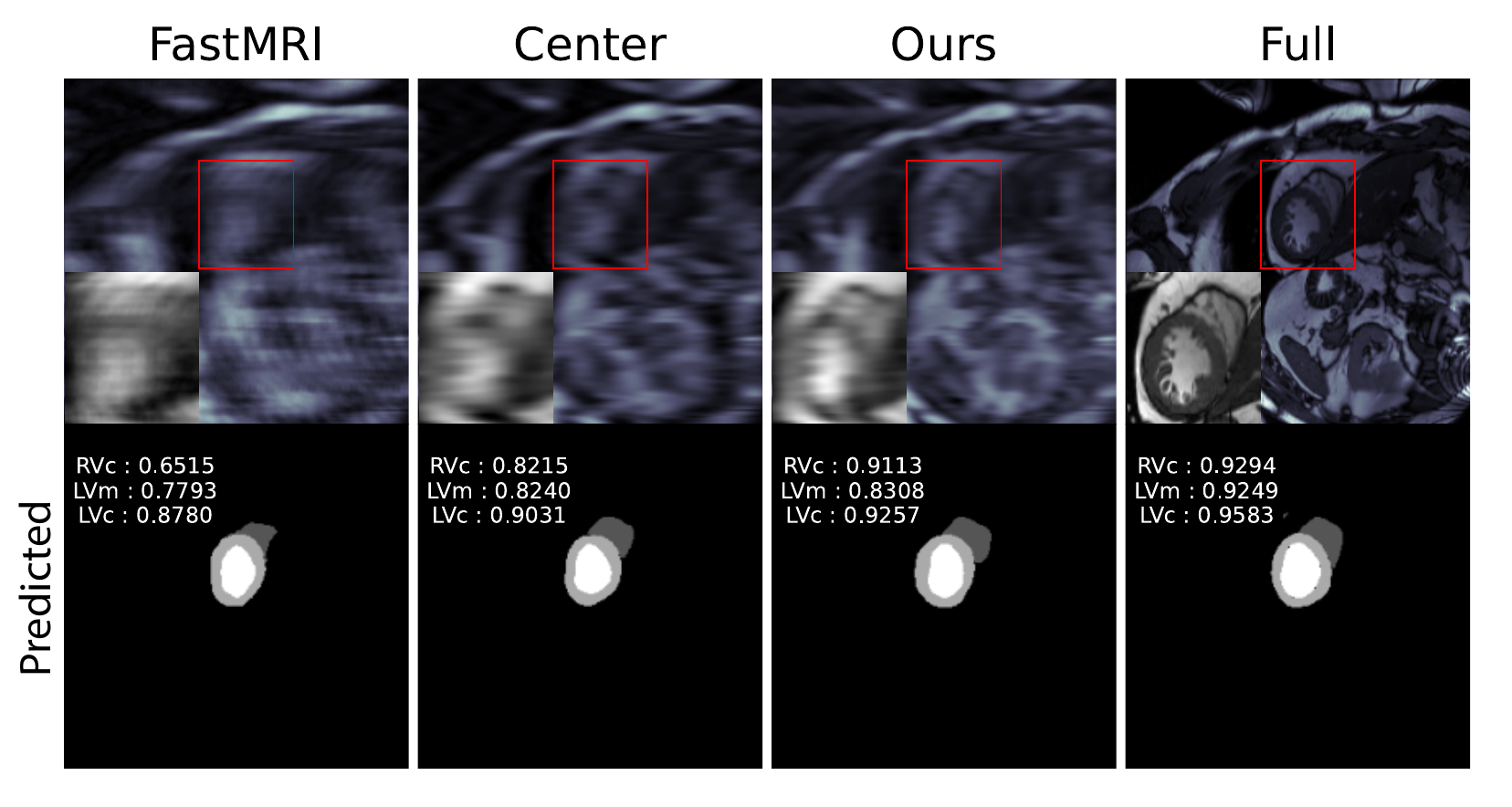}
    \caption{\textbf{Segmentation. }Fine-tuned U-Net undersampling segmentation example on ACDC at $\times$16 acceleration: left and right ventricular cavity (LVc \sqboxEmpty{black}, RVc \legendsquare{dark}) and the left ventricular myocardium (LVm \legendsquare{light}). Image with low-resolution (Center) sampling pattern has thicker right ventricle wall. Image with our undersampling pattern shows less right ventricle wall dilation.}
    \label{sf:unet_acdc_undersampled}
\end{figure}

Additional results with the ACDC dataset are reported in Fig.~\ref{sf:unet_acdc_segment} and Fig.~\ref{sf:unet_acdc_undersampled}. 


\subsection{Brain tumor segmentation examples}\label{s:brats}
\begin{figure}[!h]
    \centering
    \includegraphics[width=\linewidth]{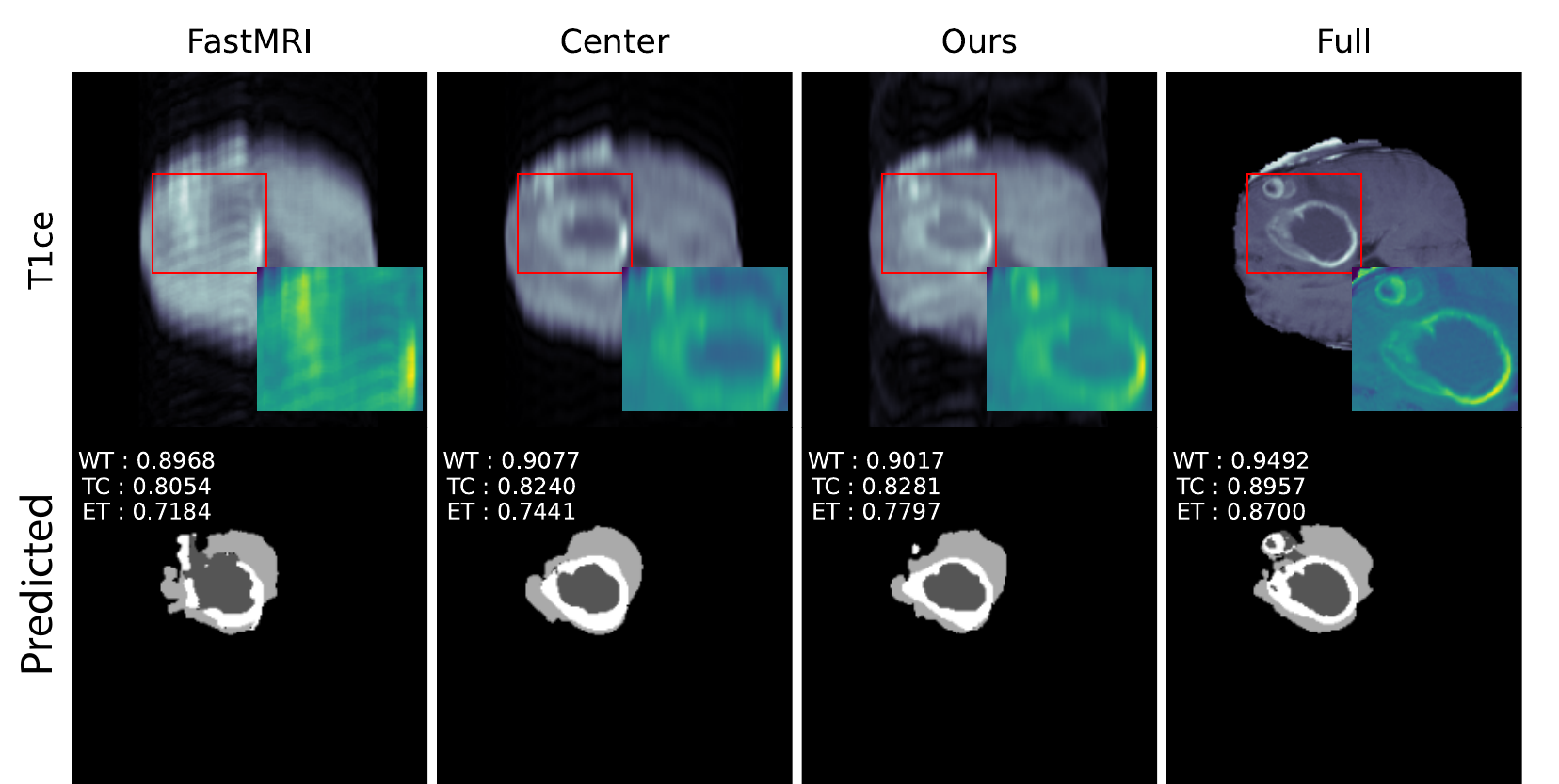}
    \caption{\textbf{Segmentation. }Fine-tuned U-Net undersampling segmentation example on BraTS at $\times$16 acceleration: whole tumor (WT \legendsquare{light}), the tumor core (TC \sqboxEmpty{black}) and the enhancing tumor (ET \legendsquare{dark}).}
    \label{sf:brats_unet}
\end{figure}

Additional results with the BraTS2020 dataset are reported in Fig.~\ref{sf:brats_unet}. The proposed IGS undersampling pattern is more sparse and provides better Dice scores than the low-resolution (Center) pattern by cost of more aliasing artifacts on T1ce modality. IGS patterns better preserve small tumor fragments nearby the larger areas, thus enabling better segmentation in terms of Dice.

\begin{figure}[!h]
    \centering
    \includegraphics[width=\linewidth]{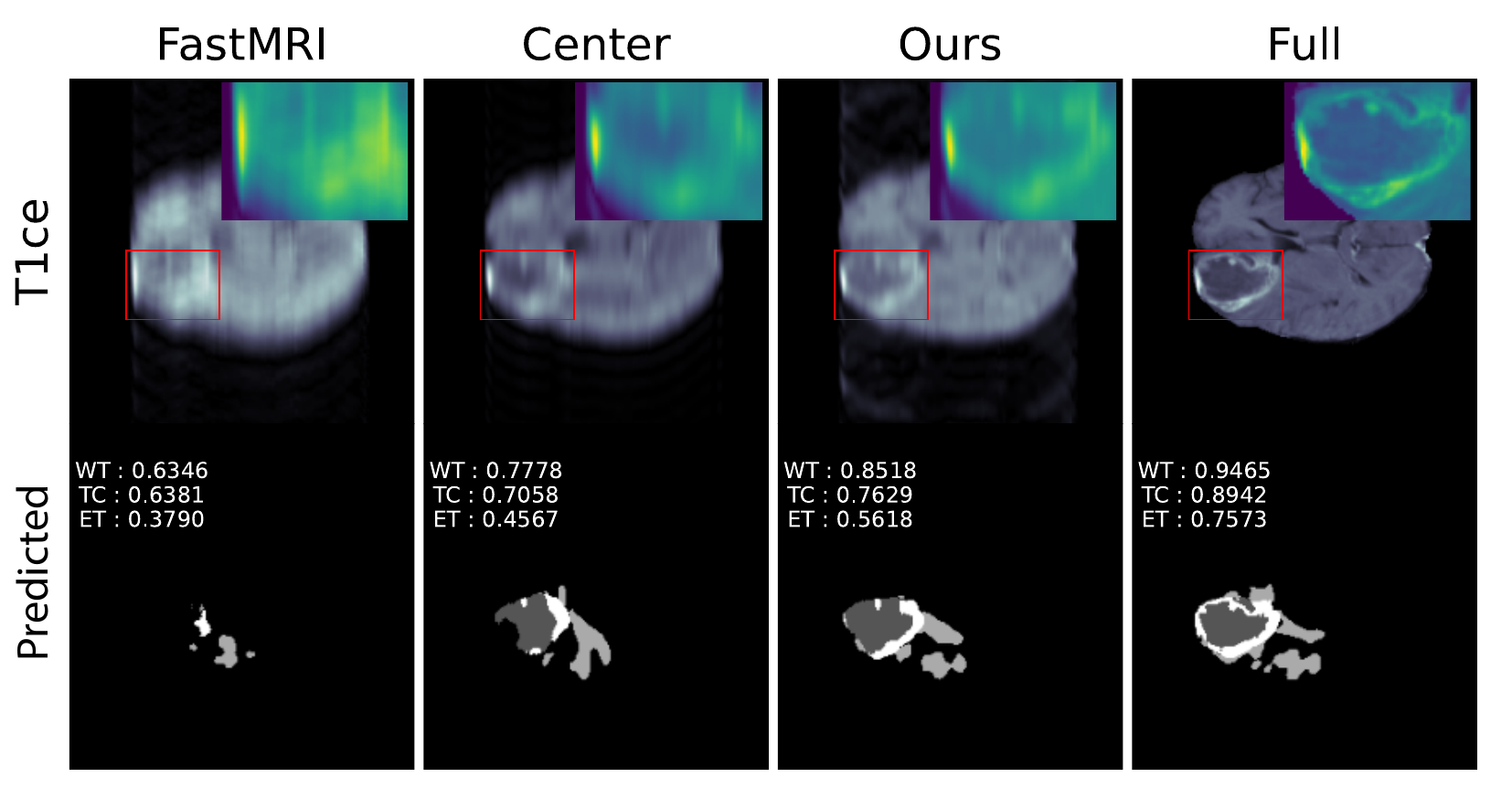}
    \caption{\textbf{Segmentation. }Fine-tuned U-Net undersampling segmentation example on BraTS at $\times$16 acceleration: whole tumor (WT \legendsquare{light}), the tumor core (TC \sqboxEmpty{black}) and the enhancing tumor (ET \legendsquare{dark})}
    \label{sf:brats_unet_2}
\end{figure}
\begin{figure}[!h]
    \centering
    \includegraphics[width=\linewidth]{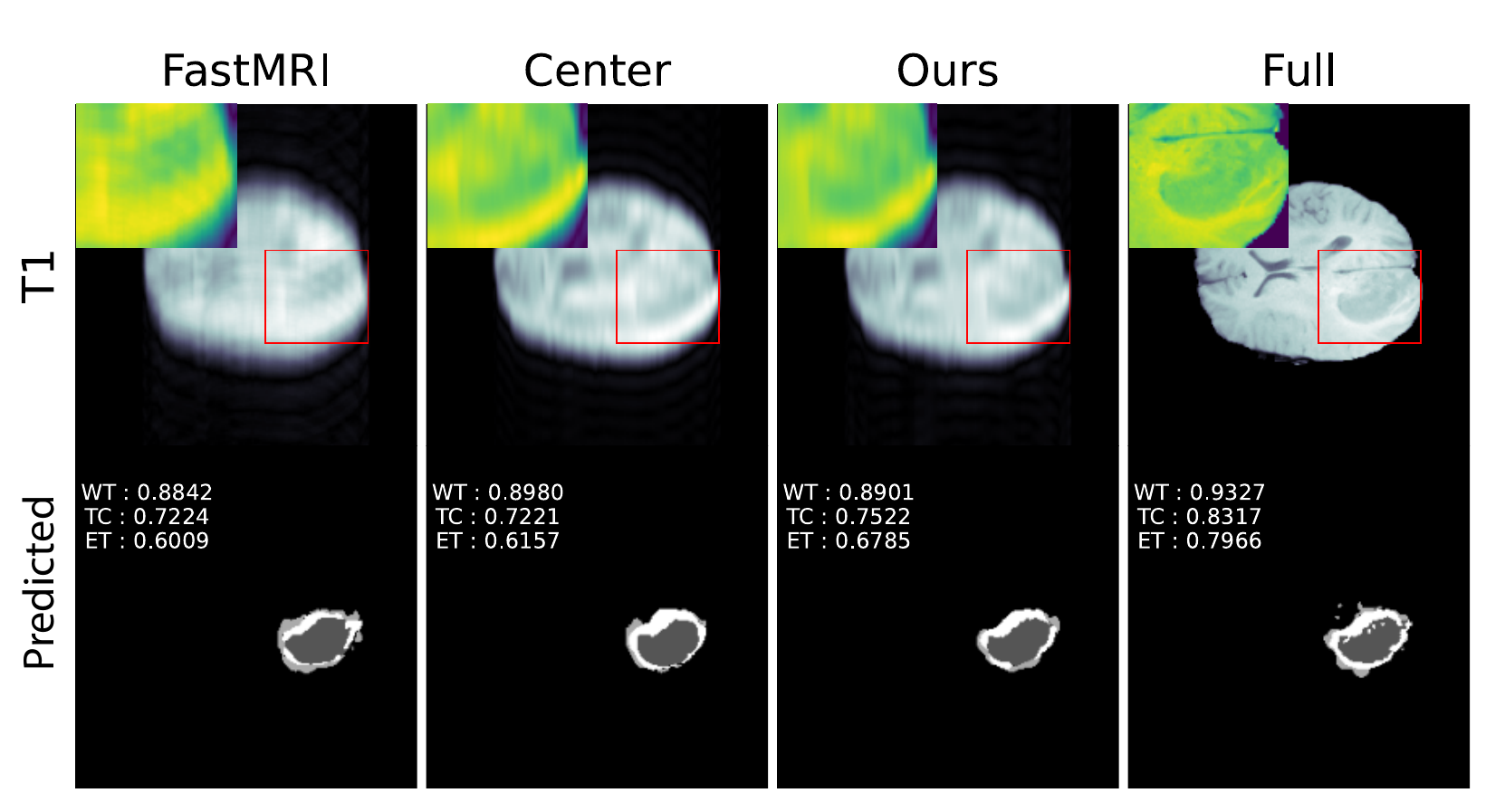}
    \caption{\textbf{Segmentation. }Fine-tuned U-Net undersampling segmentation example on BraTS at $\times$16 acceleration: whole tumor (WT \legendsquare{light}), the tumor core (TC \sqboxEmpty{black}) and the enhancing tumor (ET \legendsquare{dark})}
    \label{sf:brats_unet_3}
\end{figure}
\begin{figure}[!h]
    \centering
    \includegraphics[width=\linewidth]{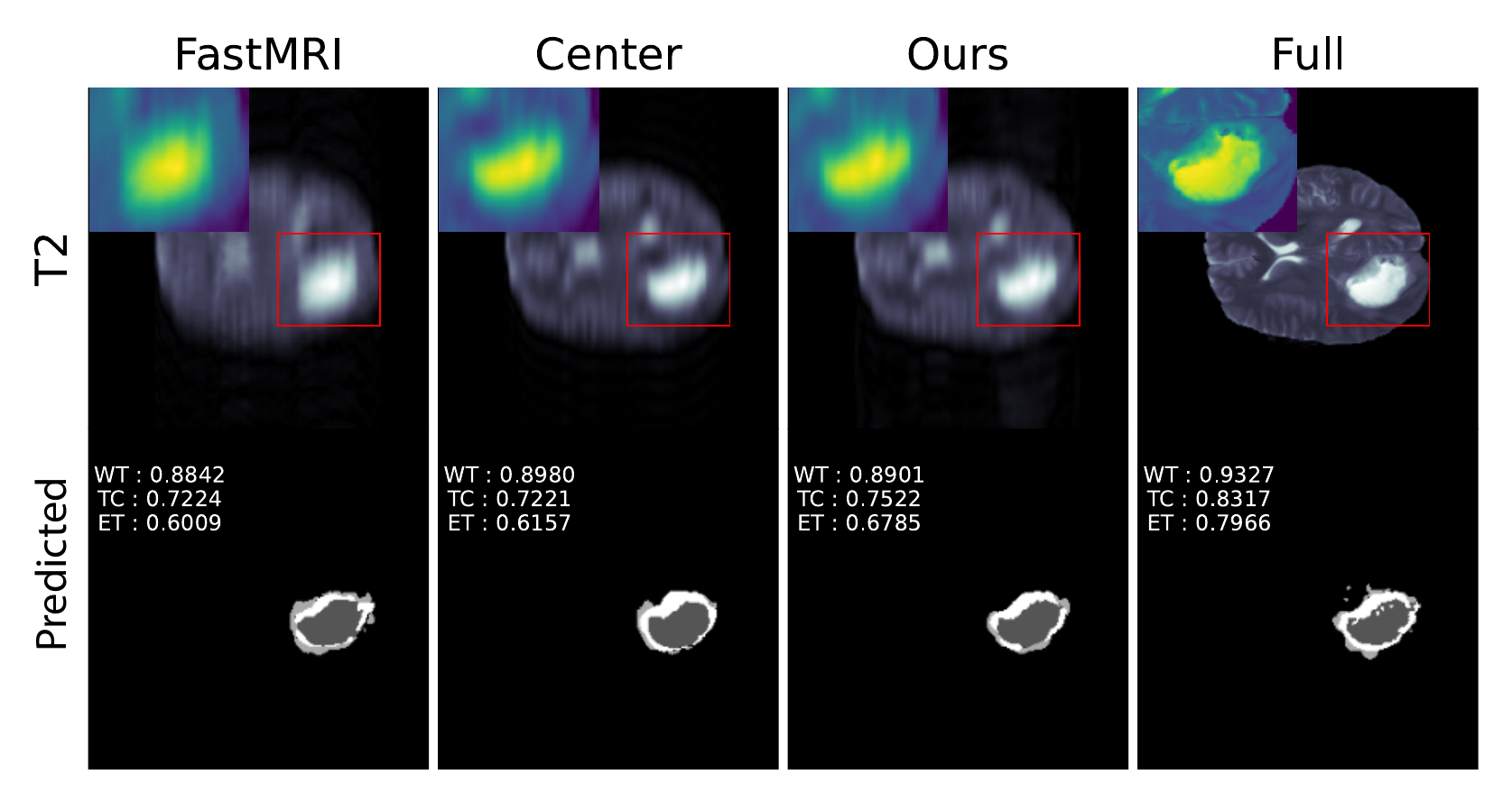}
    \caption{\textbf{Segmentation. }Fine-tuned U-Net undersampling segmentation example on BraTS at $\times$16 acceleration: whole tumor (WT \legendsquare{light}), the tumor core (TC \sqboxEmpty{black}) and the enhancing tumor (ET \legendsquare{dark})}
    \label{sf:brats_unet_4}
\end{figure}
\section{Statistics}\label{s:CV}
\begin{figure}[!h]
    \centering
    \includegraphics[width=\linewidth]{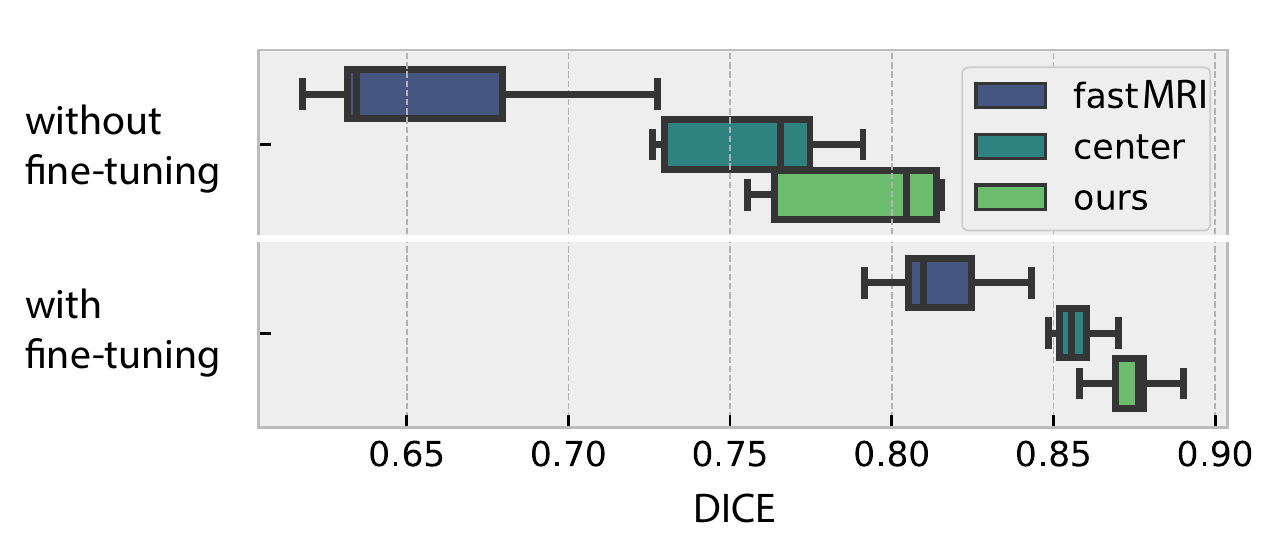}
    \caption{\textbf{Segmentation. }Dice scores on 5-fold cross-validation on ACDC. Our sampling patterns showed best Dice scores with and without fine-tuning.}
    \label{sf:5fold_acdc}
\end{figure}

\begin{figure}[!h]
    \centering
    \includegraphics[width=\linewidth]{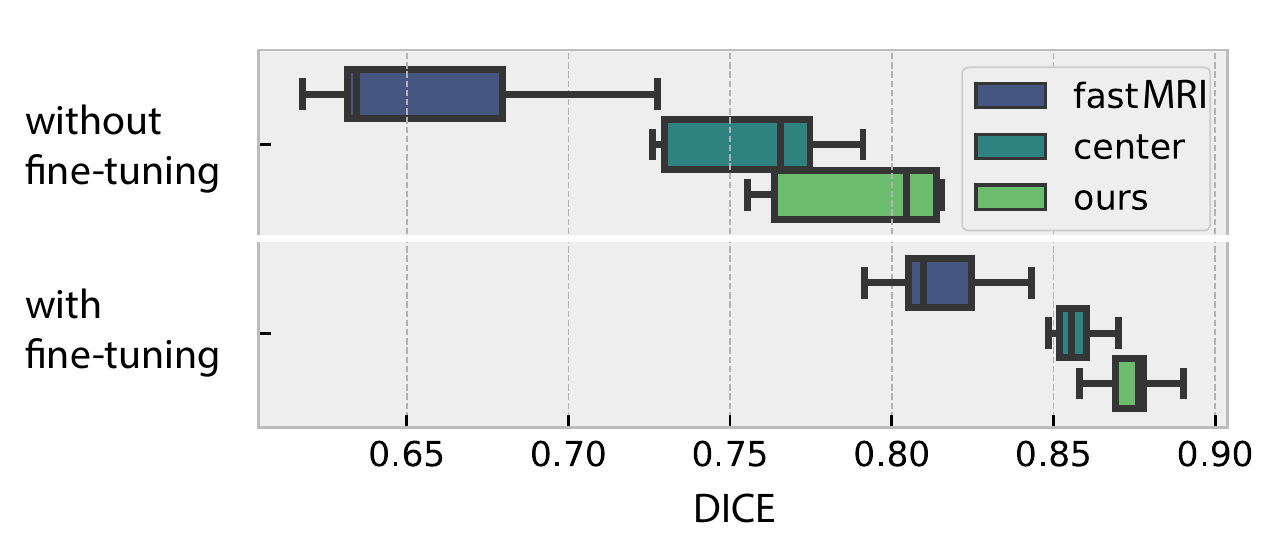}
    \caption{\textbf{Segmentation. }Dice scores on 5-fold cross-validation on BraTS dataset. Our sampling patterns showed best Dice scores with and without fine-tuning.}
    \label{sf:5fold_brats}
\end{figure}

Visualized statistics for the 5-fold cross-validation results is demonstrated in Fig.~\ref{sf:5fold_acdc} for the ACDC dataset and in Fig.~\ref{sf:5fold_brats} for the BraTS dataset.

\end{document}